\documentclass{jfm}
\usepackage{graphicx}
\usepackage{epstopdf, epsfig}
\usepackage{lipsum}
\usepackage{amsmath,amssymb}
\usepackage{amsfonts}
\usepackage{graphicx}
\usepackage{todonotes}
\usepackage{tikz}
\usepackage{color}
\usetikzlibrary{shapes}

\definecolor{MATLABcyan}{RGB}{77, 190, 238}
\definecolor{MATLABred}{RGB}{217, 83, 25}
\definecolor{MATLABgreen}{RGB}{119, 172, 48}
\definecolor{MATLAByellow}{RGB}{237, 177, 32}
\definecolor{MATLABpurple}{RGB}{126, 47, 142}
\definecolor{MATLABgrey}{RGB}{128, 128, 128}

 \newcommand{\Mgreen}{\raisebox{0.5pt}{\tikz{\node[draw=MATLABgreen,scale=0.5,circle,fill=MATLABgreen](){};}}}
 \newcommand{\Mred}{\raisebox{0.5pt}{\tikz{\node[draw=MATLABred,scale=0.5,circle,fill=MATLABred](){};}}}
\newcommand{\Myellow}{\raisebox{0.5pt}{\tikz{\node[draw=MATLAByellow,scale=0.5,circle,fill=MATLAByellow](){};}}}
\newcommand{\Mcyan}{\raisebox{0.5pt}{\tikz{\node[draw=MATLABcyan,scale=0.5,circle,fill=MATLABcyan](){};}}}
\newcommand{\Mpurple}{\raisebox{0.5pt}{\tikz{\node[draw=MATLABpurple,scale=0.5,circle,fill=MATLABpurple](){};}}}
\newcommand{\Mgrey}{\raisebox{0.5pt}{\tikz{\node[draw=MATLABgrey,scale=0.5,circle,fill=MATLABgrey,opacity=0.5](){};}}}
\newcommand{\Mgreyf}{\raisebox{0.5pt}{\tikz{\node[draw=MATLABgrey,scale=0.5,circle,fill=MATLABgrey](){};}}}
\newcommand{\Mblack}{\raisebox{0.5pt}{\tikz{\node[draw,scale=0.5,circle,fill=black](){};}}}

\shorttitle{Particle focusing dynamics in spiral ducts}
\shortauthor{R. Valani, B. Harding and Y. Stokes}

\title{Inertial particle focusing in fluid flow through spiral ducts: dynamics, tipping phenomena and particle separation}

\author{Rahil N. Valani\aff{1}
  \corresp{\email{rahil.valani@adelaide.edu.au}},
  Brendan Harding\aff{2}
 \and Yvonne M. Stokes\aff{1}}

\affiliation{\aff{1}School of Computer and Mathematical Sciences, University of Adelaide, South Australia 5005, Australia
\aff{2}School of Mathematics and Statistics, Victoria University of Wellington, Wellington 6012, New Zealand}

\begin{document}

\maketitle

\begin{abstract}
Small finite-size particles suspended in fluid flow through an enclosed curved duct can focus to points or periodic orbits in the two-dimensional duct cross-section. 
This particle focusing is due to a balance between inertial lift forces arising from axial flow and drag forces arising from cross-sectional vortices. 
The inertial particle focusing phenomenon has been exploited in various industrial and medical applications to passively separate particles by size using purely hydrodynamic effects. 
A fixed size particle in a circular duct with a uniform rectangular cross-section can have a variety of particle attractors, such as stable nodes/spirals or limit cycles, depending on the radius of curvature of the duct. 
Bifurcations occur at different radii of curvature, such as pitchfork, saddle-node and saddle-node infinite period (SNIPER), which result in variations in the location, number and nature of these particle attractors. 
By using a quasi-steady approximation, we extend the theoretical model of \citet{harding_stokes_bertozzi_2019} developed for particle dynamics in circular ducts to spiral duct geometries with slowly varying curvature, and numerically explore the particle dynamics within. Bifurcations of particle attractors with respect to radius of curvature can be traversed within spiral ducts and give rise to rich nonlinear particle dynamics and various types of tipping phenomena, such as bifurcation-induced tipping (B-tipping), rate-induced tipping (R-tipping) and a combination of both, which we explore in detail. 
We discuss implications of these unsteady dynamical behaviours for particle separation and propose novel mechanisms to separate particles by size in a non-equilibrium manner. 
\end{abstract}


\section{Introduction}

Small finite-sized particles suspended in a fluid flowing along a straight enclosed duct can migrate across streamlines due to inertia of the surrounding disturbed fluid. 
This phenomenon is known as inertial migration and the force acting on the particles to cause this is called inertial lift. 
It was first reported in experiments by \citet{SEGRE1961} who observed that small particles suspended in a straight pipe with a circular cross-section focused to an annular ring located at $0.6$ times the radius of the pipe from the centre. In general, inertial migration results in focusing of particles onto a low-dimensional attractor of the particle-fluid system, such as points or closed curves within the two-dimensional duct cross-section. 
This process is often referred to as {inertial particle focusing}. 
In straight ducts with polygonal shaped cross-sections, such as rectangles and squares, the breaking of continuous rotational symmetry of a circular cross-section typically results in multiple isolated point attractors having discrete symmetry in the cross-section~\citep{DiCarlo2009,C4LC01216J}.

The phenomenon of inertial particle focusing has found practical applications in various biomedical and industrial microfluidic technologies where passive techniques, that are minimally invasive, are desired to separate particles based on differing intrinsic properties~\citep{review2,review1}.
Current advances in inertial microfluidic particle separation technologies are primarily driven by experiments from which through trial-and-error, one can optimise inertial microfluidic devices for a specific application. 
The ability to predict and optimise particle separation for different applications based on foundational understanding of the underlying physics of inertial focusing behaviours is still a work in progress~\citep{review3}. 

Microfluidic devices aimed at exploiting inertial particle focusing to separate particles by size often utilise curved duct geometries~\citep{Liu2019}, e.g. typically consisting of a uniform rectangular cross-section extruded along a circular or spiral curve~(see figure~\ref{Fig: schematic} for a schematic). 
Within curved ducts, in addition to primary fluid flow directed through each cross-section, the fluid flow develops a pair of counter-rotating vortices within the cross-section, known as Dean vortices~\citep{Dean1927,Dean1959}. 
These cross-sectional vortices, induced by the curved geometry, break the flow symmetry across the width of the cross-section that is present in a straight duct (having the same cross-sectional shape). The interaction between Dean vortex drag and inertial lift force has three broad effects: (i) a reduction in the number of particle attractors within any given cross-section (i.e. compared to straight ducts), (ii) a pronounced dependence on particle size of the nature and/or location of particle attractors, and (iii) a modest speed-up in cross-sectional particle migration. 
These features are advantageous for particle separation, thus explaining why curved ducts are typically preferred over straight ducts for applications requiring passive particle separation.

Mathematical models of particle migration have been developed using asymptotic methods to accurately estimate the forces driving particle migration in straight ducts~\citep{hood_lee_roper_2015} and constant curvature ducts~\citep{harding_stokes_bertozzi_2019} at low flow rates. Numerical simulations using the model of \citet{harding_stokes_bertozzi_2019} have shown that for neutrally buoyant particles suspended in fluid flow through circular ducts\footnote{We use \emph{circular duct} to refer to any duct geometry where a fixed cross-section is extruded along the arc of a circle.} with rectangular cross-sections, the number, nature and location of the particle attractors varies with the radius of curvature\footnote{We utilise \emph{radius of curvature} rather than \emph{bend radius} here as it will be important to distinguish the two later when we discuss certain spiral geometries.}. 
The rich dynamical landscape, featuring a large variety of bifurcations, has been thoroughly explored by~\citet{Kyung2021} and \citet{ValaniDSTA2021,Valani2022SIADS}. 
Within a spiral duct, the local radius of curvature, and hence also the particle attractors, continuously change as a particle advances through the duct. 
A natural question which arises is: \emph{what are the implications of continuous movement of particle attractors, with the potential to cross bifurcations, on inertial particle focusing and particle separation within spiral geometries?} 
In this manuscript, we aim to address this question by systematically exploring particle focusing dynamics in spiral ducts with slowly varying curvature that have uniform rectangular cross-sections.

Due to the horizontal and vertical symmetry of the rectangular cross-section, it is common to have multiple particle attractors in circular ducts with a large radius of curvature. 
The existence of multiple attractors make the underlying dynamical system multistable. In multistable systems, there are various mechanisms, called \emph{critical transitions}, that may lead to switching of the system from one attractor to another under the influence of small perturbations. Critical transitions have become a major focus of research in climate change and ecology where the transition from one state to another might be undesirable. Sometimes the terminology used is different in different fields and these critical transitions are often referred to as \emph{tipping points} in climate science, while they are commonly known as \emph{regime shifts} in ecology~\citep{Feudel2018}. There are several possible ways in which critical transitions can occur, however there are two major types in deterministic dynamical systems~\citep{Feudel2018}. Firstly, \emph{bifurcation-induced tipping}, also known as B-tipping, takes place when a parameter of the dynamical system crosses a certain threshold value leading to a bifurcation in the system's attractor and resulting in \emph{tipping} of the dynamical state away from the previous attractor. Secondly, \emph{rate-induced tipping}, also known as R-tipping, takes place when a parameter is varied on a time scale different from that of the internal dynamics of the system; when the parameter's rate of change exceeds a critical threshold, then parts of the system are not able to successfully track an attractor causing the dynamical state to \emph{tip} away from that attractor towards another. In spiral ducts, the continuously changing radius of curvature can result in the emergence of both types of tipping phenomena as a result of bifurcations and continuous movement of particle attractors. 
This results in a complex dynamical landscape which we explore in detail by extending our studies \citep{harding_stokes_bertozzi_2019,Valani2022SIADS} of particle focusing and dynamics in circular ducts  to spiral ducts with slowly varying curvature.
We systematically explore the different dynamical behaviours and tipping phenomena as the system parameters are varied. 
Based on our observations, we also discuss how the observed dynamical features can be exploited for non-equilibrium particle separation in spiral ducts. 

The paper is organised as follows. In section~\ref{sec: formalism} we present the theoretical model and rationalise our choice of an Archimedean-like spiral duct geometry. We then briefly review the bifurcations that take place with radius of curvature in constant curvature ducts in section~\ref{sec: bifurcations}. In section~\ref{sec: spiral without bif} we present the results of particle dynamics in spiral ducts when no bifurcations are present, while in section~\ref{sec: spiral with bif} we explore spiral ducts with bifurcations that result in rich particle dynamics and tipping phenomena. We briefly explore the effects of flow rate on the particle dynamics in section~\ref{sec: flow rate}. In section~\ref{sec: particle separation}, we discuss novel non-equilibrium particle separation mechanisms. Our conclusions are given in section~\ref{sec: conclusions}.

\section{Theoretical Formalism}\label{sec: formalism}

\subsection{Mathematical model}\label{sec: model}

\begin{figure}
\centering
\includegraphics[width=0.95\columnwidth]{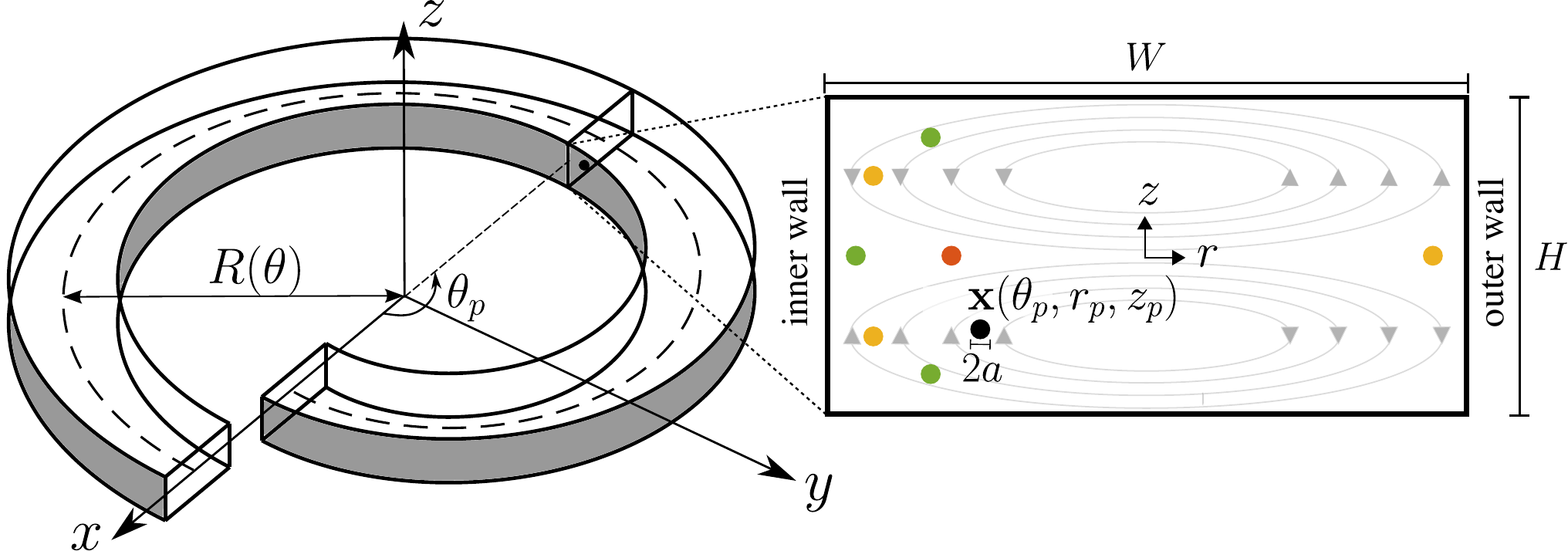}
\caption{Schematic of the theoretical setup. A particle of radius $a$ with centre located at $\mathbf{x}_p=\mathbf{x}(\theta_p,r_p,z_p)$ is suspended in an incompressible fluid flow through a Archimedean spiral duct with changing bend radius, $R(\theta)$, and having a uniform rectangular cross-section of width $W$ and height $H$. The enlarged view of the cross-section illustrates the local cross-sectional $(r,z)$ co-ordinate system and the approximate streamlines of the secondary flow (grey closed curves) induced by the curvature of the duct. The edge labelled ``inner wall" is the side closer to origin $(x,y,z)=(0,0,0)$ while the edge labelled ``outer wall" is the side further away from the origin. The black filled circle~(\protect\Mblack) denotes the particle location while the coloured circles denote the particle equilibria:  unstable node in red~(\protect\Mred), stable nodes (point attractors) in green~(\protect\Mgreen) and saddle points in yellow~(\protect\Myellow).}
\label{Fig: schematic}
\end{figure}

\citet{harding_stokes_bertozzi_2019} developed an asymptotic model to calculate the forces acting on a neutrally buoyant spherical particle suspended in fluid flow through a constant curvature duct at sufficiently small flow rates. 
By balancing these leading order forces with viscous Stokes' drag they constructed a first order quasi-steady model for the trajectory of the particle. In this work, we adapt their model and apply it to spiral ducts with slowly varying curvature. By \emph{slowly varying curvature} we mean that the background fluid flow through any cross-section of the spiral duct does not differ significantly from that through a circular duct with identical curvature. We expect the quasi-steady approximation of inertial migration to remain valid for such spiral ducts when approximating the background flow accordingly. 
The quality of this approximation will be demonstrated in section~\ref{sec: justification}. Below we briefly review the model of \citet{harding_stokes_bertozzi_2019}; further details may be obtained from the paper. 

As shown in figure~\ref{Fig: schematic}, consider a neutrally buoyant spherical particle of radius $a$ suspended in fluid flowing through a spiral duct with a uniform rectangular cross-section. 
This specific example illustrates an Archimedean spiral for which the \emph{bend radius}, $R(\theta)$, varies with the azimuthal angle, $\theta$. 
The rectangular cross-section has width $W$ and height $H$.
In a circular duct (with constant bend radius $R(\theta)=R$), and in the absence of the particle, a constant pressure gradient along the duct drives a steady incompressible fluid flow~\citep{Dean1927,Dean1959} which has been studied extensively for ducts with a rectangular cross-section~\citep{winters_1987,Yamamoto_2004,Harding2018}. In addition to the axial component of the fluid flow, the curvature induces a secondary flow within the cross-section consisting of a counter-rotating vortex pair. The presence of a particle in the duct disturbs this background flow and one can define disturbance pressure and velocity fields by taking the difference between the fields in the presence and absence of the particle. The resulting dimensionless dynamical equations for the disturbance pressure field $q$ and disturbance fluid velocity field $\mathbf{v}$ in a rotating frame (moving with the particle) are given by
\begin{align}\label{Eq: NS}
- \nabla q + \nabla^2\mathbf{v}&=\text{Re}_p \left( \left(\mathbf{v}+\bar{\mathbf{u}}-\Theta \left(\mathbf{e}_{z}\times \mathbf{x}\right)\right)\cdot \nabla \mathbf{v}+\mathbf{v}\cdot\nabla \bar{\mathbf{u}} +\Theta\left(\mathbf{e}_z\times\mathbf{v}\right)\right) \:\:\: \text{on}\:\:\mathbf{x} \in \mathcal{F}, \nonumber\\
\nabla \cdot \mathbf{v} &= 0 \:\:\: \text{on}\:\:\mathbf{x} \in \mathcal{F},\\ \nonumber
\mathbf{v}&=\mathbf{0} \:\:\: \text{on}\:\:\mathbf{x} \in \partial \mathcal{D},\\ \nonumber
\mathbf{v}&=-\bar{\mathbf{u}}+\Theta\left(\mathbf{e}_z\times\mathbf{x}\right)+\mathbf{\Omega}_p\times\left(\mathbf{x}-\mathbf{x}_p\right) \:\:\: \text{on}\:\:\mathbf{x} \in \partial \mathcal{F} \setminus \partial \mathcal{D}.
\end{align}
Here, $\bar{\mathbf{u}}$ is the background fluid flow velocity in the absence of the particle, $\mathbf{\Omega}_p$ is the particle spin and $\mathbf{e}_{z}$ is the unit vector in the vertical $z$ direction; all quantities are in a reference frame which is rotating about the $z$ axis at a rate $\Theta:=\partial \theta_p/\partial t$ which is assumed to be (nearly) constant. The domain $\mathcal{D}$ denotes the interior of the duct, $\partial \mathcal{D}$ denotes the boundaries of the duct, $\mathcal{F}:=\{\mathbf{x}\in\mathcal{D}:|\mathbf{x}-\mathbf{x}_p|\geq 1\}$ is the dimensionless fluid domain in the presence of the particle and $\partial \mathcal{F} \setminus \partial \mathcal{D}=\{\mathbf{x}:|\mathbf{x}-\mathbf{x}_p|=1\}$ is the surface (boundary) of the particle. The dimensionless parameter $\text{Re}_p$ in equation~\eqref{Eq: NS} is the particle Reynolds number defined as
$\text{Re}_p=\text{Re}\,(a/l)^2$,
{where} $\text{Re}=\rho U_m l/\mu$ is the flow Reynolds number, $\rho$ {is} the particle/fluid density, $\mu$ is the fluid's dynamic viscosity, $l=\text{min}\{W,H\}$ is the characteristic length scale associated with the duct cross-section and $U_m$ is a characteristic velocity of the background fluid flow. The variables in equation~\eqref{Eq: NS} have been scaled with physical parameters as follows: the particle radius $a$ for $\mathbf{x}$, $l/U_m$ for time $t$, $U_m a/l$ for the velocities $\mathbf{v}$ and $\bar{\mathbf{u}}$, $U_m/l$ for the particle spin $\mathbf{\Omega}_p$ and the angular velocity $\Theta$, and $\mu U_m/l$ for the disturbance pressure $q$. 

The disturbed fluid flow exerts force and torque on the spherical particle. The dimensionless force (scaled by $\rho U_m^2 a^4/l^2$) and torque (scaled by $\rho U_m^2 a^5/l^2$) acting on the particle in the rotating frame are given by
\begin{align}\label{eq: force particle}
\mathbf{F}=&-\frac{4\pi}{3}\Theta^2 \left(\mathbf{e}_z\times\left( \mathbf{e}_z\times \mathbf{x}_p \right) \right)+ \int_{|\mathbf{x}-\mathbf{x}_p|<1} \bar{\mathbf{u}}\cdot\nabla\bar{\mathbf{u}}\,\text{d}V \\ \nonumber
&+\frac{1}{\text{Re}_p} \int_{|\mathbf{x}-\mathbf{x}_p|=1} (-\mathbf{n})\cdot(-q\mathbf{I}+\nabla\mathbf{v}+\nabla\mathbf{v}^T)\,\text{d}S,
\end{align}
\begin{align}\label{eq: torque particle}
\:\:\:\:\:\:\:\:\:\:\:\:\:\:\:\:\:\:\:\:\:\mathbf{T}=&-\frac{8\pi}{15}\Theta \left(\mathbf{e}_z\times\mathbf{\Omega}_p \right)+ \int_{|\mathbf{x}-\mathbf{x}_p|<1} (\mathbf{x}-\mathbf{x}_p)\times(\bar{\mathbf{u}}\cdot\nabla\bar{\mathbf{u}})\,\text{d}V \\ \nonumber
&+\frac{1}{\text{Re}_p} \int_{|\mathbf{x}-\mathbf{x}_p|=1} (\mathbf{x}-\mathbf{x}_p)\times((-\mathbf{n})\cdot(-q\mathbf{I}+\nabla\mathbf{v}+\nabla\mathbf{v}^T))\,\text{d}S.  
\end{align}
Taking the particle Reynolds number as a small parameter and performing a perturbation expansion
of the disturbance flow in powers of $\text{Re}_p$ we get
\begin{align*}
\mathbf{v} &= \mathbf{v}_0+\text{Re}_p \mathbf{v}_1+{O}(\text{Re}^2_p), \\ \nonumber
q &= q_0+\text{Re}_p q_1+{O}(\text{Re}^2_p).
\end{align*}
Substituting into equation \eqref{Eq: NS} gives us a zeroth order system {for} $q_0,\mathbf{v}_0$, and a first order system {for} $q_1,\mathbf{v}_1$. The zeroth order system captures all  the non-zero boundary conditions and the first order system captures the most significant inertial contribution to the complete equations. 
Substituting the perturbation expansions in equation~\eqref{eq: force particle} results in the following equation for the leading-order cross-sectional force on the particle
\begin{equation}\label{Eq: cross-sec force}
    \mathbf{F}_s=(\mathbf{e}_r\cdot(\text{Re}^{-1}_p\mathbf{F}_{-1,s}+\mathbf{F}_0))\,\mathbf{e}_r+(\mathbf{e}_z\cdot(\text{Re}^{-1}_p\mathbf{F}_{-1,s}+\mathbf{F}_0))\,\mathbf{e}_z.
\end{equation}
Here, $\mathbf{F}_{-1}$ is the leading order force consisting of drag from the background flow, with the subscript `s' denoting the secondary (cross-sectional) component, while $\mathbf{F}_0$ is the leading order contribution of the inertial lift force.
Balancing the cross-sectional force in equation~\eqref{Eq: cross-sec force} with viscous Stokes' drag, we construct a first order quasi-steady model for the trajectory of the particle giving the following dynamical equations of motion:
\begin{equation*}
    \frac{\text{d} r_p}{\text{d} t}=-\text{Re}_p\frac{F_{s,r}}{C_r},\:\:\:
    \frac{\text{d} z_p}{\text{d} t}=-\text{Re}_p\frac{F_{s,z}}{C_z}\:\:\:\text{and}\:\:\:\frac{\text{d} \theta_p}{\text{d} t}=\frac{\bar{{u}}_a}{R/a+r_p},
\end{equation*}
where $\bar{{u}}_a$ is the axial component of the background fluid flow, $F_{s,r}=\mathbf{F}_s\cdot\mathbf{e}_r$ and $F_{s,z}=\mathbf{F}_s\cdot\mathbf{e}_z$ are the radial and the vertical components of the cross-sectional force, respectively, with corresponding drag coefficients $C_r$ and $C_z$ that vary with the particle's position in the cross-section. 
Thus, this model estimates the particle velocities within a circular duct
with each of the forces implicitly depending on the radius of curvature (which coincides with the bend radius $R$) in addition to the particle radius $a$, each measured relative to the duct height. 

We are able to extend this quasi-steady model to spiral ducts with slowly varying curvature by considering $R$ as a function of $\theta$ to describe the local radius of curvature (which must be distinguished from any notion of bend radius in non-circular duct geometries, see section~\ref{sec: justification} for more details). The radius of curvature is the appropriate quantity to determine local fluid behaviour as it is a well-defined geometrical construct which applies to any curve, whereas the notion of bend radius relies on having a well-defined origin (or other point of reference) which doesn't apply to generalised curves. Thus, while it is convenient to describe circular and Archimedean spiral ducts via a ``bend radius", we need to consider the radius of curvature when modelling the fluid.

The circular duct model is solved numerically using a finite element method to compute the forces acting on the particle. We refer the reader to \citet{harding_stokes_bertozzi_2019} for additional details on the numerical framework. Once the relevant forces have been pre-computed at numerous sample points over the cross-section (and for numerous system parameter values), smooth interpolants of $C_r$, $C_z$, $F_{s,r}$, $F_{s,z}$ are constructed and the particle dynamics are simulated using the MATLAB solver ode45.


For the results presented in this manuscript, we use the following non-dimensional scales for the system variables: dimensionless radius of curvature (or bend radius, as appropriate) $\tilde{R}=2R/H$, dimensionless particle size $\tilde{a}=2a/H$, dimensionless time $\tilde{t}=U_m t/H$, and cross-sectional co-ordinates $\tilde{r}=2r/H$ and $\tilde{z}=2z/H$.

\subsection{Spiral duct geometry}

Our illustrations thus far have utilised an Archimedean spiral, since these are readily recognised and understood.
However, to extend the circular duct model to spiral ducts, it is convenient to instead consider an Archimedean-like spiral duct whose centreline is described by the parametric equation:
\begin{equation}\label{Archimedean-like spiral}
    \mathbf{\tilde{r}}_I(\varphi)=[\tilde{x}(\varphi),\tilde{y}(\varphi)]=\tilde{R}(\varphi)[\cos(\varphi),\sin(\varphi)]+\beta[-\sin(\varphi),\cos(\varphi)-1],
\end{equation}
where 
\begin{align}
\tilde{R}(\varphi)&=\tilde{R}_{\text{start}}+\beta\varphi\,, \label{Rc equation}\\
\beta &= \frac{\tilde{R}_{\text{end}}-\tilde{R}_{\text{start}}}{2\pi N_{\text{turns}}}\,, \label{beta def}
\end{align}
with $\tilde{R}_{\text{start}},\tilde{R}_{\text{end}}$ denoting the radius of curvature at the start and end of the spiral duct, respectively, and $N_{\text{turns}}$ denoting the number of turns. 
This curve corresponds to the involute of a circle (having radius $\beta$) and, although visually similar, differs in several ways from a traditional Archimedean spiral described by
\begin{equation}\label{Archimedean spiral}
    \mathbf{\tilde{r}}_A(\theta)=[\tilde{x}(\theta),\tilde{y}(\theta)]=\tilde{R}(\theta)[\cos(\theta),\sin(\theta)],
\end{equation}
with $\tilde{R}(\theta)$ as in \eqref{Rc equation}.

For the traditional Archimedean spiral of \eqref{Archimedean spiral}, the radial distance from the origin, often referred to as the \emph{bend radius}, varies linearly with respect to $\theta$. 
Accordingly, $\tilde{R}_{\text{start}},\tilde{R}_{\text{end}},\tilde{R}(\theta)$ should be interpreted as \emph{bend radii} in the context of the Archimedean spiral.
However, it is important to note that the \emph{bend radius} of an Archimedean spiral does not equal its \emph{radius of curvature}, and it is the latter which is fundamental to describing/modelling the local fluid flow.
In contrast, for the involute spiral of equation~\eqref{Archimedean-like spiral} it is the \emph{radius of curvature} that varies linearly with respect to $\varphi$ and  $\tilde{R}_{\text{start}},\tilde{R}_{\text{end}},\tilde{R}(\varphi)$ can be interpreted as \emph{radii of curvature} at the appropriate locations.
The trade-off is that $\varphi$ no longer describes the polar angle.

There are a number of additional differences between the two curves worth noting.
The involute spiral has a closed-form arclength parametrisation whereas the Archimedean spiral does not.
The principal unit normal of the involute spiral is $\mathbf{N}_I(\varphi)=[-\cos(\varphi),-\sin(\phi)]$ and is directed from $\tilde{\mathbf{r}}_I(\varphi)$ to the point $\beta[-\sin(\varphi),\cos(\varphi)-1]$.
The principal unit normal for the Archimedean spiral is more complex and, importantly, is \emph{not} directed from $\mathbf{\tilde{r}}_A(\theta)$ towards the origin.
These differences between the two curves are subtle when $\beta$ is small, but they become important when considering how to update the value of $\tilde{R}$ as a particle traverses a spiral duct having positive width. 
The simplest way one might describe the side walls in both spirals is to replace $\tilde{R}(\cdot)$ with $\tilde{R}(\cdot)\pm\tilde{W}/2$ in each of \eqref{Archimedean-like spiral} and \eqref{Archimedean spiral}.
For the involute spiral this leads to a consistent width of $\tilde{W}$ as measured with respect to the principal unit normal of $\tilde{\mathbf{r}}_I(\varphi)$, but the same cannot be said for the Archimedean spiral.
Moreover, when considering how to update $\tilde{R}$ as a particle traverses the involute spiral we have
\begin{equation}
\frac{d}{d\tilde{t}}\tilde{R}(\varphi)=\beta\frac{d\varphi}{d\tilde{t}}=\beta\frac{d\varphi}{ds_I}\frac{ds_I}{d\tilde{t}}\,,
\end{equation}
where $s_I(\varphi)$ is the arclength along the involute spiral.
The term $ds_I/d\tilde{t}$ is simply the velocity of the particle down the main axis (i.e. aligned with the tangent vector $\hat{\mathbf{T}}_I(\varphi)=[-\sin(\varphi),\cos(\varphi)]$). 
For $d\varphi/ds_I$ we note that $ds_I/d\varphi=\tilde{R}(\varphi)+\tilde{r}_p$ where the addition of $\tilde{r}_p$ accounts for the radial coordinate of the particle within the cross-section.
For the Archimedean spiral, we similarly have
\begin{equation}
\frac{d}{d\tilde{t}}\tilde{R}(\theta)=\beta\frac{d\theta}{d\tilde{t}}=\beta\frac{d\theta}{ds_A}\frac{ds_A}{d\tilde{t}}\,,
\end{equation}
where $s_A(\varphi)$ is the arclength along the Archimedean spiral.
The term $ds_A/d\tilde{t}$ is again the particle velocity down the main axis, however one must recognise that the tangent vector is given by the more complex expression
\begin{equation*}
\hat{\mathbf{T}}_A(\theta)=\frac{\tilde{R}(\theta)[-\sin(\theta),\cos(\theta)]+\beta[\cos(\theta),\sin(\theta)]}{\sqrt{\tilde{R}(\theta)^2+\beta^2}}\,.
\end{equation*}
Moreover, for $d\theta/ds_A$ we note that $ds_A/d\theta=\sqrt{\tilde{R}(\theta)^2+\beta^2}+\tilde{r}_p$.
Upon accounting for these subtleties, one must still determine the radius of curvature of the Archimedean spiral to ensure the correct migration forces are sampled from our circular duct model.
All together, these differences make the involute spiral much better suited to extending our curved duct model of inertial particle migration. 
We will use $\tilde{\mathbf{r}}(\varphi)=\tilde{\mathbf{r}}_I(\varphi)$ and $\tilde{R}=\tilde{R}(\varphi)$ going forwards, where $\tilde{R},\tilde{R}_{\text{start}},\tilde{R}_{\text{end}}$ should always be interpreted as radii of curvature (with the exception of Appendix~\ref{A} where we compare results with an Archimedean spiral).

When designing a spiral duct, upper and lower bounds on the number of turns, ${N}_\text{turns}$, is determined by both geometrical and physical/practical constraints. 
From a purely geometrical point of view, the maximum number of turns, $N_{\max}$, is constrained by the fact that the spiral duct, having non-zero width along the spiral curve, should not intersect itself. 
This means that after one turn of the spiral, $\tilde{R}(\varphi)$ should change at least by the width of the cross-section resulting in the constraint $2\pi\beta>\tilde{W}$, or equivalently $N_{\max}<|\tilde{R}_{\text{end}}-\tilde{R}_{\text{start}}|/\tilde{W}$ , where $\tilde{W}=W/H$ is the aspect ratio of the rectangular cross-section. 
In addition to this geometrical constraint, there may be additional  practical experimental/design constraints that further limit the maximum number of turns. 
For example, the required pressure difference to drive fluid flow through the duct increases with the length of the duct and there may be an upper limit on the pressure difference to preserve the structural integrity of a microfluidic device. 
The minimum number of turns $N_{\min}$ is constrained by two factors: 
(i) the maximum change in radius of curvature that can be considered reasonable in our slowly varying curvature model, 
and (ii) the time/length required for the particles (initially randomly distributed in the cross-section) 
to focus to
(sufficiently close to) their particle attractors. 
Both of these considerations are also influenced by the flow rate at which the device is expected to operate at.

\subsection{Numerical test for the slowly varying curvature approximation}\label{sec: justification}

Steady flow through an Archimedean spiral duct having a rectangular cross-section with large aspect ratio was considered in \citet{Harding2018}. (Non-rectangular cross-sections were also considered but in the context of this work only the rectangular cross-sections are of interest.) 
An approximation was developed via a regular perturbation expansion with respect to both the curvature parameter and the height-to-width ratio.
It was found that the leading order solution depended on the local curvature of the spiral but not its rate of change. 
First order corrections with respect to both perturbation parameters were also considered, but these too did not introduce any non-trivial dependence on the rate of change of the spiral curvature. 
This provides strong evidence that spiral duct flow does not differ significantly from circular 
duct flow (when using the appropriate curvature parameter at any given point in the spiral).
More specifically, it suggests that non-trivial dependence of the flow with respect to our $\beta$ parameter only occurs at order $O(\epsilon^2)$ or smaller (relative to the magnitude of each leading order component).

To test this further we conducted a numerical simulation of fluid flow through a relatively short spiral duct having a somewhat large value of $\beta$.
Specifically, we considered a setup with $\tilde{R}_{\mathrm{start}}=100$, $\tilde{R}_{\mathrm{end}}=20$, $N_{\mathrm{turns}}=1/2$, $W=4$ and $H=2$, i.e. such that $|\beta|=80/\pi\approx 25.5$.
We generated a structured mesh of this domain consisting of roughly 780,000 tetrahedra and then solved the Navier--Stokes equations using the finite element method with Taylor--Hood elements (utilising the FEniCS computing platform \citep{LoggEtal2012,AlnaesEtal2015}).
At different cross-sections we then compared the velocity field with that which is obtained by solving the equations of steady flow through a circular 
duct \citep{harding_2019,Harding2022}. 
At the cross-section of the spiral duct for which the radius of curvature is $50$, and with a flow Reynolds number of $100$, we measured a relative $L_2$ difference in the axial, radial and vertical velocity components of the circular 
and spiral duct flows as $3.3\%$, $5.5\%$ and $3.1\%$ respectively.
The error decreased for smaller Reynolds numbers, for example with $\mathrm{Re}=50$ we observe relative differences of $0.53\%$, $0.99\%$ and $2.0\%$, respectively.

In a second computation we generated 
a structured mesh of a spiral duct with $\tilde{R}_{\mathrm{start}}=50$, $\tilde{R}_{\mathrm{end}}=300$, $N_{\mathrm{turns}}=1$ and other parameters as above, such that now $|\beta|=125/\pi\approx40.8$.
The same number of tetrahedra were used in this case.
At the cross-section for which the radius of curvature is $200$, and again with a flow Reynolds number of $100$, we measured a relative $L_2$ difference in the three velocity components as $0.25\%$, $2.6\%$ and $0.46\%$.
In this instance the difference did not decrease significantly when decreasing the Reynolds number, which suggests that the dominant error is due to the mesh resolution.
A similar observation was made for other cross-sections (which aren't too close or far from the ends where there are inlet/outlet effects).
A precise analysis of the difference between circular 
and spiral duct flow is the subject of ongoing work, but together these results indicate that $\beta=O(10)$ can be considered small enough that the spiral geometry is sufficiently slowly varying that the fluid flow through any given cross-section can be well approximated by circular duct flow with appropriately chosen curvature parameter, for Reynolds numbers up to $O(100)$.


\section{Bifurcations with respect to radius of curvature}\label{sec: bifurcations}

\begin{figure}
\centering
\includegraphics[width=0.95\columnwidth]{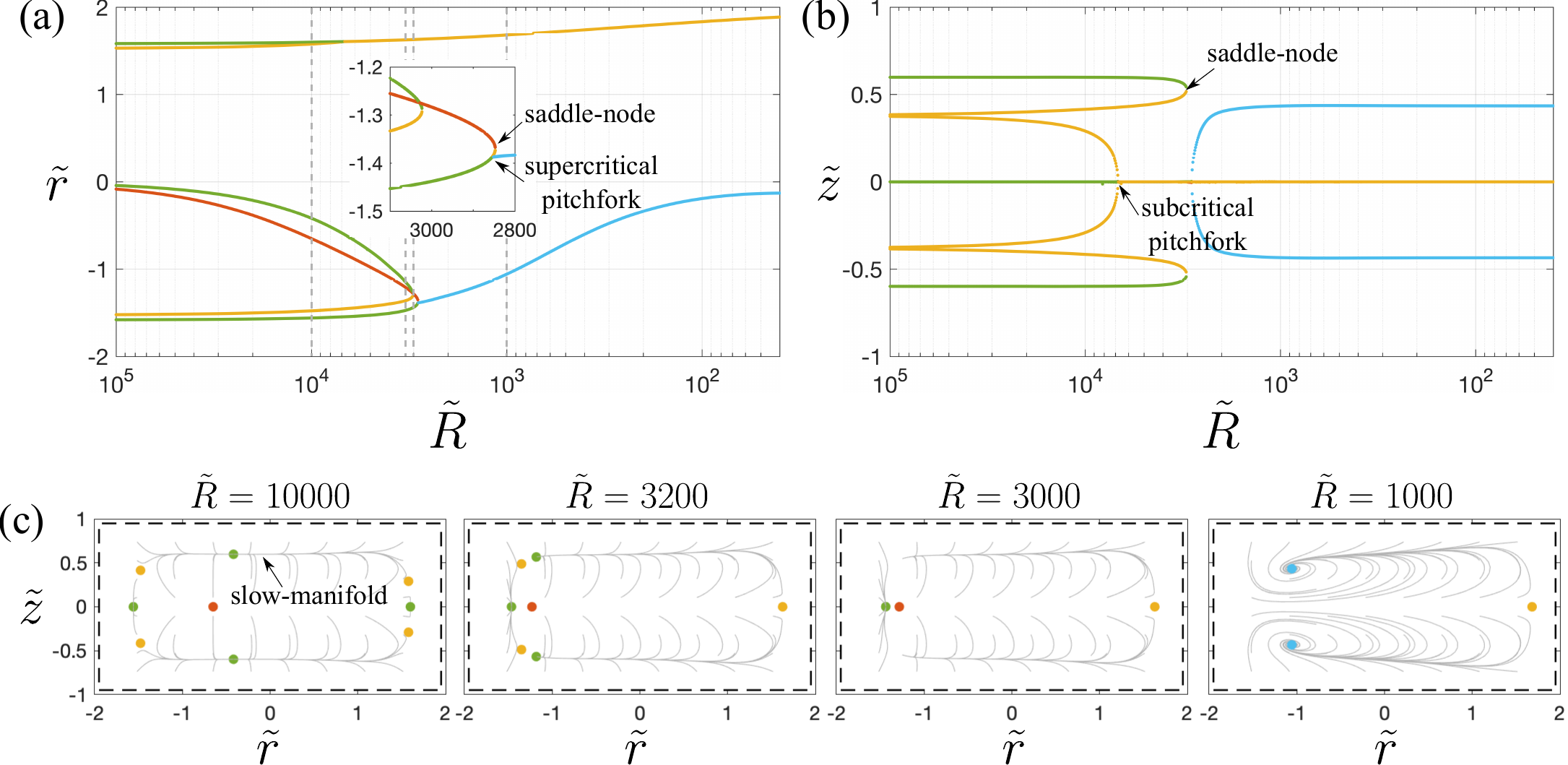}
\caption{A typical bifurcation diagram showing equilibria for a particle of dimensionless radius $\tilde{a}=2a/H=0.05$ in constant curvature ducts with a $2\times1$ rectangular cross-section. (a) Horizontal dimensionless location $\tilde{r}=2r/H$ and (b) vertical dimensionless location $\tilde{z}=2z/H$ of the particle equilibria as a function of the dimensionless radius of curvature $\tilde{R}=2R/H$. The inset in panel (a) shows details of bifurcations near $\tilde{R}\approx 3000$. (c) Snapshots showing the particle equilibria in the two-dimensional cross-section at various $\tilde{R}$ values. These radii of curvature correspond to the vertical grey dashed lines in panel (a). For each panel, the size of the circle corresponds to particle size and the colour denotes the type of equilibria:  unstable nodes in red~(\protect\Mred), stable nodes/spirals (point attractor) in green~(\protect\Mgreen)/cyan~(\protect\Mcyan) and saddle points in yellow~(\protect\Myellow). The  grey  curves  in panel (c) illustrate some  trajectories  of  particles  within  the  cross-section while the dashed rectangle indicates the location of the centre of the particle for which it will touch the walls of the duct.}
\label{Fig: bifurcations}
\end{figure}


We start by briefly reviewing the effect of radius of curvature on particle equilibria in circular duct geometries. We refer the interested reader to \citet{harding_stokes_bertozzi_2019} and \citet{Valani2022SIADS} for a more detailed exploration of how various system parameter such as radius of curvature, cross-sectional geometry and particle size, change the particle equilibria. Here we provide an illustrative example of bifurcations with respect to changes in radius of curvature which 
will provide a 
foundation for the rest of the paper. 

Figures~\ref{Fig: bifurcations}(a) and (b) show bifurcation diagrams depicting the radial ($\tilde{r}$) and vertical ($\tilde{z}$) co-ordinates of particle equilibria (i.e. locations where the net force on the particle is zero), respectively, as a function of the radius of curvature $\tilde{R}$ for a small particle of radius $\tilde{a}=0.05$ in a rectangular $2\times1$ cross-section. The particle equilibria in the two-dimensional cross-section for $\tilde{R}=10000,3200,3000$ and $1000$ are shown in panel (c). We see that at relatively large radii of curvature, many particle equilibria are found in the cross-section, for example an unstable node near the centre of the duct, saddle-points near the corners and stable nodes (point attractors) near the edges of the rectangular cross-section. A universal dynamical feature observed at relatively large radii of curvature is a slowing of migration along heteroclinic orbits which connect saddle points to stable nodes. These connected unstable/stable manifolds of saddle points/stable nodes effectively constitute a \emph{slow-manifold}.
Several bifurcations take place as the radius of curvature is decreased. Firstly, near $\tilde{R}\approx7000$, a subcritical pitchfork bifurcation takes place near the outer wall where a stable node and two saddle points merge into a single saddle point. As the radius of curvature is further decreased, a pair of saddle-node bifurcations take place near $\tilde{R}\approx3000$ where saddle points near the inner wall annihilate with stable nodes near the top and bottom edges. Further decreasing the radius of curvature leads to two rapid bifurcations (see the inset of figure~\ref{Fig: bifurcations}(a)). The stable node on the horizontal centreline ($\tilde{z}=0$) near the inner wall first undergoes a supercritical pitchfork bifurcation producing two stable nodes in the vertical direction with a saddle point in between. The saddle point then merges with the unstable node in a saddle-node bifurcation while the two stable nodes transition to stable spirals. No further bifurcations take place as the radius of curvature is decreased even further, but the pair of stable spirals asymptotically migrate horizontally towards the vertical centreline ($\tilde{r}=0$) of the duct as $\tilde{R}\xrightarrow{}\tilde{W}$.

For different particle sizes in the range $0<\tilde{a}\leq0.2$, similar bifurcations take place but at different radii of curvature $\tilde{R}$. \citet{harding_stokes_bertozzi_2019} showed that for circular ducts with rectangular cross-sections, the variations in particle equilibria with $\tilde{R}$ for different sized particles can be captured reasonably well by a single dimensionless ratio $\kappa=4/(\tilde{a}^3 \tilde{R})$, which describes the scaling of secondary flow drag relative to the inertial lift force. The bifurcation curves for different $\tilde{a}$ when plotted as a function of $\kappa$ collapse reasonably well onto a single curve.

\section{Spiral ducts with no bifurcations in particle equilibria}\label{sec: spiral without bif}

We start by analysing the dynamics of neutrally buoyant particles in spiral ducts when there are no bifurcations of particle equilibria over the entire range of radii of curvature within the spiral. 
Despite an absence of bifurcations, there can still be non-trivial dynamical effects due to the motion of stable nodes/spirals (point attractors) or due to changes in the size and dynamics along limit cycles ($1$D attractors). 
We treat cases where there is a vertically symmetric pair of stable nodes/spirals or limit cycles ($\pm \tilde{z}$) in the duct cross-section as a single particle attractor~(with respect to the horizontal $\tilde{r}$ direction). 
We explore the case of a single point attractor in section~\ref{sec: no bif single}, a single limit cycle pair in section~\ref{sec: no bif lim cyc}, while the presence of multiple particle attractors in the horizontal direction of the cross-section is explored in section~\ref{sec: no bif multiple}.   

\subsection{Spiral ducts with a single point attractor}\label{sec: no bif single}

We start by analysing the dynamics of particles in spiral ducts where there is a single particle attractor. In ducts with rectangular cross-sections, this regime arises at smaller radii of curvature where secondary drag forces dominate inertial lift forces and one obtains either: (i) a single stable node near the inner wall of the curve duct (e.g. see $\tilde{R}=3000$ in figure~\ref{Fig: bifurcations}(c)), or (ii) a vertically symmetric pair of stable nodes/spirals (e.g. see $\tilde{R}=1000$ in figure~\ref{Fig: bifurcations}(c)).  

\begin{figure}
\centering
\includegraphics[width=0.9\columnwidth]{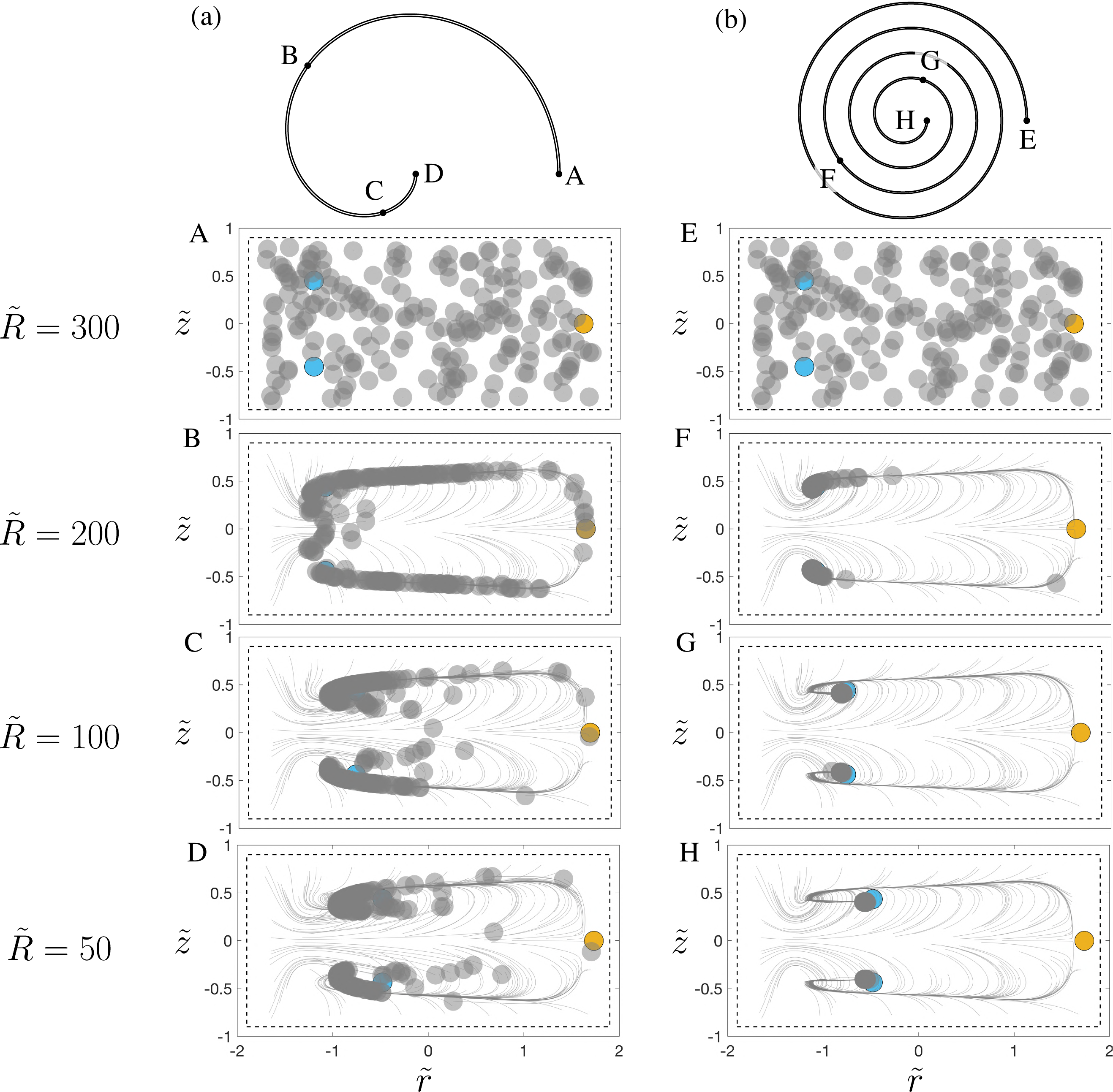}
\caption{Particle focusing dynamics in an in-spiral duct with a $2\times1$ rectangular cross-section that has a vertically symmetric pair of point attractors. The system parameters are $R_{\text{start}}=300$, $R_{\text{end}}=50$, $\text{Re}=50$ and $\tilde{a}=0.10$. Snapshots of the cross-section are shown at (A,E) $\tilde{R}=300$, (B,F) $200$, (C,G) $100$ and (D,H) $50$, for column (a) $N_{\text{turns}}=1$ and column (b) $N_{\text{turns}}=4$, respectively. The coloured circles denote the type of particle equilibria with cyan~(\protect\Mcyan) for stable spirals (point attractors) and yellow~(\protect\Myellow) for a saddle point. The grey circles~(\protect\Mgrey) denote the particle positions while the grey curves denote their trajectories. If the centre of a particle lies on the dashed rectangle, 
it will touch at least one wall of the duct.}
\label{Fig: stable spiral single}
\end{figure}

A typical example of particle focusing dynamics in an in-spiral duct (i.e. the particle travels in the direction of decreasing radius of curvature) consisting of a vertically symmetric pair of stable spirals at the same horizontal $\tilde{r}$ location, is shown in figure~\ref{Fig: stable spiral single}. Here we compare particle focusing dynamics for spirals having the same starting ($\tilde{R}_{\text{start}}$) and ending ($\tilde{R}_{\text{end}}$) radii of curvature but a different number of turns ($N_{\text{turns}}$). Note that even in the absence of bifurcations, the particle equilibria are not static throughout the spiral ducts; they move in response to the changing radius of curvature of each cross-section. We observe that for the spiral having less turns, the combination of insufficient duct length and continuous movement of the point attractors inhibits complete particle focusing and a spread in the particle distribution  is observed at the end of the spiral duct~(see figure~\ref{Fig: stable spiral single}(a)). Alternatively, with more turns, one observes complete focusing of particles at the end of the spiral duct~(see figure~\ref{Fig: stable spiral single}(b)). 
For cross-sections G through to H, the focused particle cluster tracks the moving attractors closely and there is a small ``lag'' between the centre of the focused clusters and the centre of the attractors. Hence, in a typical spiral duct with a single horizontal attractor and sufficient turns, the particle focusing is similar to circular ducts whose radius of curvature coincides with the local radius of curvature at the end of the spiral; in both cases the particles focus in close proximity to the single attractor whose location is determined by the radius of curvature at the end of the spiral. Nevertheless, spiral duct geometries have an advantage over circular ducts since one can increase the duct length by having multiple non-intersecting turns (which is not practically possible in circular ducts~\footnote{Circular ducts can be wrapped around in the form of a helix to obtain multiple turns while avoiding self-intersection, although the effects of torsion may become important and our present model may not be applicable.}). This becomes particularly useful at smaller radius of curvature where one turn of a circular duct is not sufficient for particle focusing~(e.g. see figure~\ref{Fig: PS1}(a)). 

\subsection{Spiral ducts with a single limit cycle pair}\label{sec: no bif lim cyc}

\begin{figure}
\centering
\includegraphics[width=0.9\columnwidth]{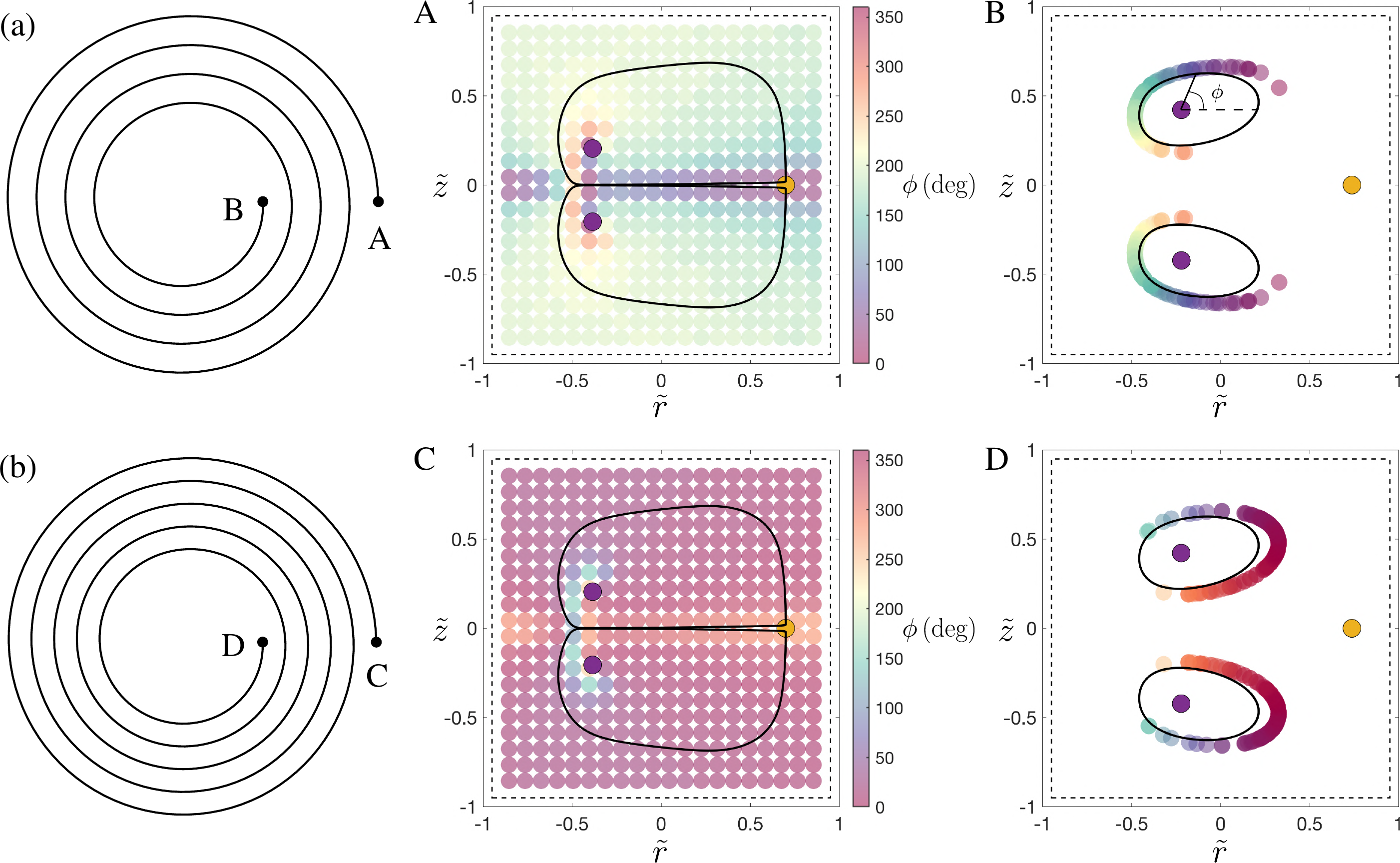}
\caption{Particle focusing dynamics in a spiral duct with square cross-section having a vertically symmetric pair of limit-cycle attractors. Rows (a,b) show particle dynamics for in-spiral ducts having (a) $N_{\text{turns}}=4$ (see Supplemental Video $1$) and (b) $N_{\text{turns}}=5$ (see Supplemental Video $2$). Cross-sectional images at the (A,C) start and (B,D) end of the spiral show particle equilibria (dark coloured circles with purple~(\protect\Mpurple) for unstable spirals and yellow~(\protect\Myellow) for saddle points) along with particle positions (light coloured circles) whose colour is based on the phase angle $\phi$ occupied by the particles along the limit cycle (black curves) at the end of the spiral. If the centre of a particle lies on the dashed square, 
it will touch at least one wall of the duct. The system parameters are fixed to $R_{\text{start}}=1250$, $R_{\text{end}}=500$, $\tilde{a}=0.05$ and $\text{Re}=50$.}
\label{Fig: stable limit single}
\end{figure}

Next, we consider an in-spiral duct with square cross-section having a single $1$D attractor in the horizontal direction, as shown in figure~\ref{Fig: stable limit single}. Specifically, we have a vertically symmetric pair of stable limit cycles (black curves) which persists throughout the entire length of the spiral. However, the decreasing radius of curvature throughout the in-spiral duct leads to a continuous change in the size of the limit cycle and the dynamics along it. The limit cycle is large in extent near the inlet of the spiral and shrinks progressively along the spiral, while the period to traverse the limit cycle simultaneously decreases. 
We make an interesting observation regarding the phase of the limit cycle occupied by particles based on their initial location in the cross-section. We find that most particles that don't start near the unstable spiral equilibria or along the horizontal centreline of the duct, aggregate around the same phase of the limit cycle~(see Supplemental Video $1$). 
One of the key factors for this particle aggregation is that initially, the majority of particles are ``squeezed" along the horizontal centreline where their dynamics slow down. This results in aggregation of particles as they approach the stable limit cycles.
This phenomenon is robust with respect to the number of turns, as can be seen in figures~\ref{Fig: stable limit single}(a) and (b) respectively~(compare Supplemental Videos $1$ and $2$), which allows the phase occupied by the majority of particles within the final cross-section to be manipulated by modifying the number of turns accordingly.

\subsection{Spiral ducts with multiple particle attractors}\label{sec: no bif multiple}

\begin{figure}
\centering
\includegraphics[width=\columnwidth]{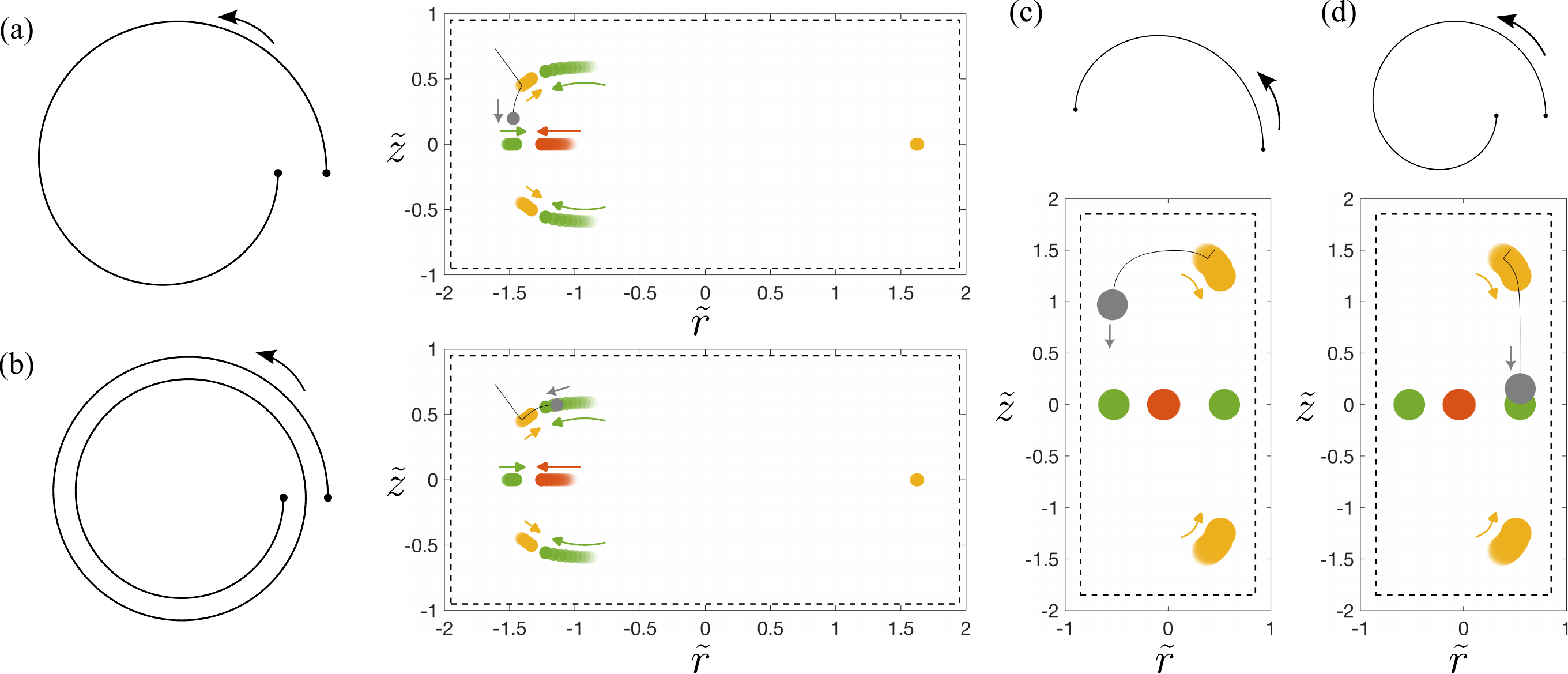}
\caption{Rate-induced tipping (R-tipping) for small particles of (a,b) radius $\tilde{a}=0.05$ in an in-spiral duct with $\tilde{R}_{\text{start}}=4500$, $\tilde{R}_{\text{end}}=3100$, $\text{Re}=50$ and a $2\times1$ rectangular cross-section. Particles which start near a saddle point that separates the basins of attraction of adjacent stable node attractors, can either fall in the basin of attraction of the left-most stable node for (a) $N_{\text{turns}}=1$, or one of the stable node pair to the right of this for (b) $N_{\text{turns}}=2$. Similar behaviour is observed in (c,d) an in-spiral duct with $\tilde{R}_{\text{start}}=1000$, $\tilde{R}_{\text{end}}=500$, $\text{Re}=50$ and a $1\times2$ rectangular cross-section for larger particles of radius $\tilde{a}=0.15$ with (c) $N_{\text{turns}}=0.5$ and (d) $N_{\text{turns}}=1$. The cross-sections of the ducts show particle equilibria (coloured circles) and their motion as the radius of curvature changes through the spiral (coloured arrows) with unstable nodes in red~(\protect\Mred), stable nodes (point attractors) in green~(\protect\Mgreen) and saddle points in yellow~(\protect\Myellow). The particle location at the end of each spiral duct (grey filled circle -~\protect\Mgreyf), its motion (grey arrows) and trajectory (black curve) are also shown. If the centre of a particle lies on the dashed rectangle 
it will touch at least one wall of the duct.}
\label{Fig: RIT}
\end{figure}

Multiple point attractors with different horizontal locations can coexist in circular ducts with rectangular cross-sections. When this occurs, the cross-sectional domain can be partitioned based on the basin of attraction for each point attractor. 
This allows one to readily identify where a particle will converge based on its initial location.
However, when spiral ducts contain multiple point attractors throughout (with no bifurcations), both the point attractors and the basins of attraction move in response to the continuously changing radius of curvature. 
This can lead to some particles finding themselves in a different basin of attraction to the one they started in and thereby induces non-trivial focusing effects.

Figures~\ref{Fig: RIT}(a) and (b) compare a particle's trajectory in two spiral duct geometries having different numbers of turns.
The duct has a $2\times1$ rectangular cross-section and the spiral is such that multiple point attractors persist throughout. Two horizontally distinct stable nodes are present, one on the horizontal centreline near the inner wall, and a vertically symmetric pair of stable nodes a bit further away from the inner wall. Two saddle points between the three stable node attractors separate their respective basins of attraction. In figure~\ref{Fig: RIT}(a) we show the trajectory of a particle starting out near the top inner (left) corner of the cross-section, specifically within the basin of the upper stable node (with respect to the starting radius of curvature) but near the boundary of that basin. As the particle is carried through the spiral duct, within the cross-section it is initially attracted towards the upper saddle point within a neighbourhood of its stable manifold. By the time the particle reaches its shortest distance to the saddle point, it finds itself in the basin of the left most stable node due to the upward motion of the saddle point~(see yellow arrows). 
Conversely, with more turns of the spiral as shown in figure~\ref{Fig: RIT}(b), a particle starting from the same initial position remains in the basin of attraction of the upper most stable node throughout the spiral. 
Thus, this particle-fluid system exhibits rate-induced tipping (R-tipping) where, depending on the rate at which the radius of curvature changes (e.g. via manipulating $N_{\text{turns}}$), particles starting near a basin boundary can approach a stable node with a basin of attraction different from the one they started in.
Similarly, rate-induced tipping occurs in a spiral duct with a tall $1\times2$ rectangular cross-section as shown in figures~\ref{Fig: RIT}(c) and (d).

We note that since the saddle point equilibria, which separate the basins of attraction, don't move significantly between the starting and ending cross-sections of the spirals shown in figure~\ref{Fig: RIT}, the R-tipping phenomenon only affects a small fraction of particles with initial positions near the basin boundaries. Hence, we do not expect R-tipping by itself to have major consequences on inertial particle focusing and separation. However, if R-tipping occurs in combination with bifurcations in particle equilibria, then significant changes in focusing behaviour can occur, as will be discussed in section~\ref{sec: spiral bif multiple}.

\section{Spiral ducts with bifurcations in particle equilibria}\label{sec: spiral with bif}

Bifurcations of particle equilibria can take place as the radius of curvature changes along a spiral duct. 
When bifurcations take place at a critical radius of curvature (depending also on the particle size and cross-sectional shape), it results in abrupt changes to particle dynamics and focusing behaviour. Moreover, since the dynamics of the system undergo a topological change, the direction in which bifurcations are traversed, i.e. in-spiral or out-spiral, leads to significantly different focusing behaviours at the end of the spiral. In this section, we explore different types of particle dynamics and focusing behaviours in spiral ducts where bifurcations occur. We start by exploring particle dynamics in spiral ducts with bifurcations at a single radius of curvature in section~\ref{sec: spiral bif single}, followed by a similar exploration in spiral ducts with bifurcations at multiple radii of curvature in section~\ref{sec: spiral bif multiple}.

\subsection{Spiral ducts with bifurcations at a single radius of curvature}\label{sec: spiral bif single}

\begin{figure}
\centering
\includegraphics[width=\columnwidth]{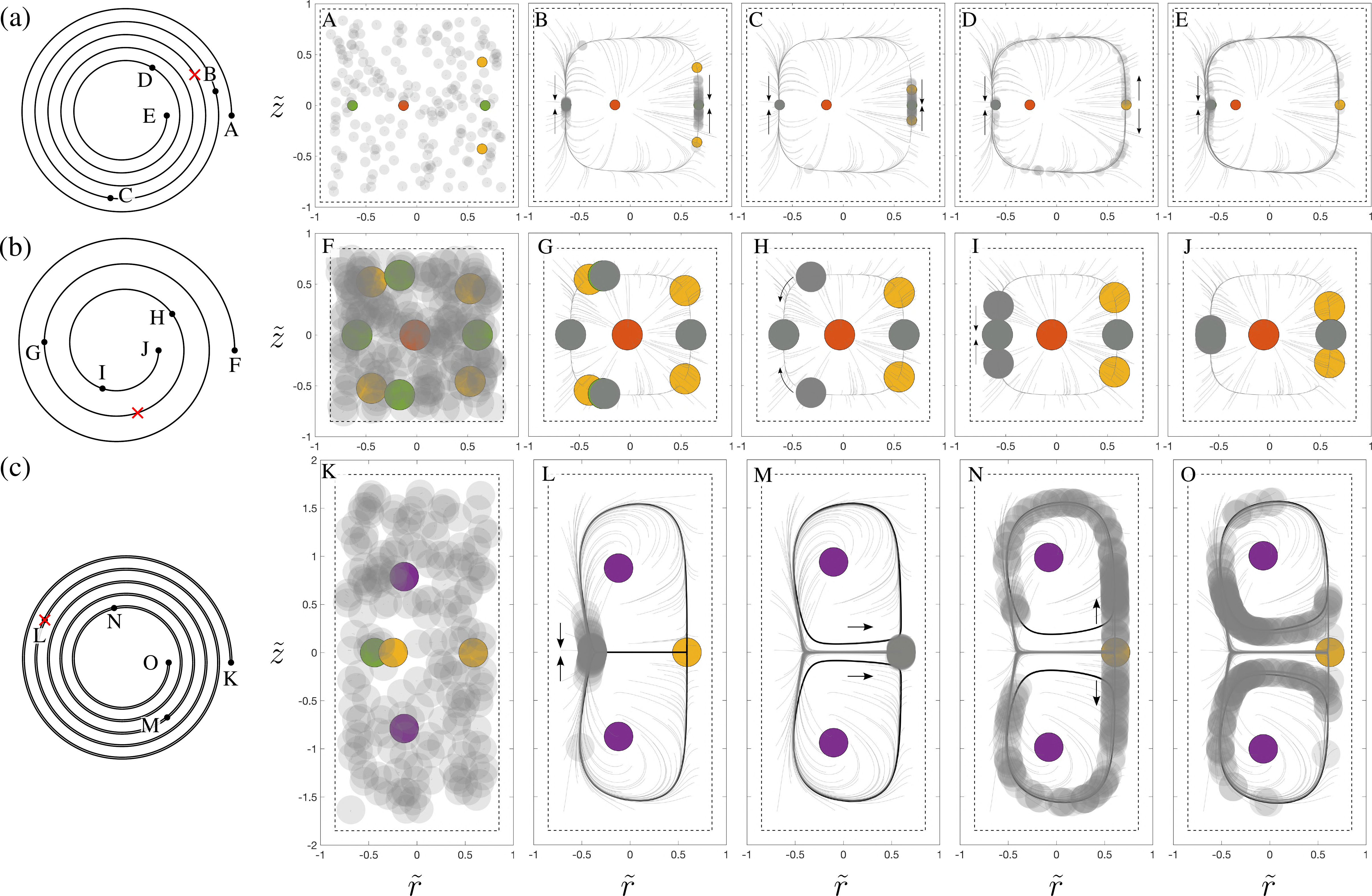}
\caption{Bifurcation-induced tipping (B-tipping) at a flow Reynolds number of $\text{Re}=50$. Row\,(a): an in-spiral duct with $\tilde{R}_{\text{start}}=4000$, $\tilde{R}_{\text{end}}=1600$, $N_{\text{turns}}=5$ and a square cross-section, in which particles of radius $\tilde{a}=0.05$ undergo B-tipping following a subcritical pitchfork bifurcation at $\tilde{R}_{cr}\approx 3000$; snapshots (A-E) are shown at $\tilde{R}=4000, 3500, 3170, 2000, 1600$. Row~(b): an in-spiral duct with $\tilde{R}_{\text{start}}=1500$, $\tilde{R}_{\text{end}}=500$, $N_{\text{turns}}=3$ and a square cross-section, in which particles of radius $\tilde{a}=0.15$ undergo B-tipping following a pair of saddle-node bifurcations at $\tilde{R}_{cr}\approx 900$; snapshots (F-J) are shown at $\tilde{R}=1500, 1000, 800, 600, 500$. Row~(c): an in-spiral duct with $\tilde{R}_{\text{start}}=120$, $\tilde{R}_{\text{end}}=50$, $N_{\text{turns}}=5$ and a rectangular $1\times2$ cross-section, in which particles of radius $\tilde{a}=0.15$ undergo B-tipping following a saddle-node infinite period (SNIPER) bifurcation at $\tilde{R}_{cr}\approx100$; snapshots (K-O) are shown at $\tilde{R}=120, 100, 80, 60, 50$. The cross-sectional images show the particle equilibria as coloured circles with unstable nodes in red~(\protect\Mred), stable nodes (point attractor) in green~(\protect\Mgreen), saddle points in yellow~(\protect\Myellow) and unstable spirals in purple~(\protect\Mpurple). Stable limit cycles are shown as black curves. Particle locations (grey circles -~\protect\Mgrey), their motion (black arrows) and trajectories (grey curves) are also shown. If the centre of a particle lies on the dashed square/rectangle 
it will touch at least one wall of the duct. Red cross~({\color{red}{$\boldsymbol{\times}$}}) on spirals denote the location of the bifurcation. 
}
\label{Fig: BIT}
\end{figure}

We first investigate particle dynamics in spiral ducts where bifurcations take place at a single critical radius of curvature, $\tilde{R}_\text{cr}$. Figure~\ref{Fig: BIT}(a) shows the focusing dynamics in an in-spiral duct with a square cross-section containing small particles ($\tilde{a}=0.05$). At the starting radius of curvature, there are two stable node attractors in the cross-section located on the horizontal centreline near the inner and outer walls. Randomly distributed particles in the cross-section initially approach these two point attractors along the slow-manifold. However, at $\tilde{R}_\text{cr}\approx 3000$, the stable node near the outer wall takes part in a subcritical pitchfork bifurcation which ultimately makes this location unstable. All of the particles which initially migrated towards the stable node near the outer wall, which is replaced with a saddle after the bifurcation occurs, now migrate towards the stable node located near the inner wall (along heteroclinic orbits which form a slow manifold). Hence, we observe bifurcation-induced tipping (also known as B-tipping) where a focused particle cluster at an attractor \emph{tips} due to a bifurcation that makes this attractor unstable. A second example of B-tipping due to saddle-node bifurcations in an in-spiral duct with square cross-section for larger particles ($\tilde{a}=0.15$) is depicted in figure~\ref{Fig: BIT}(b). In both of these examples, since the bifurcations occur on a slow-manifold, the particle migration is slow after the B-tipping takes place.

Bifurcation-induced tipping can also cause a cluster of particles which first focus to a point attractor, to ultimately migrate towards a $1$D limit cycle attractor. Figure~\ref{Fig: BIT}(c) illustrates this for 
particles of size $\tilde{a}=0.15$ in an in-spiral duct with $1\times2$ rectangular cross-section. At the starting radius of curvature, the system has only one attractor, a point attractor on the horizontal centreline near the inner wall to which all particles are initially attracted. However, this stable node vanishes at $\tilde{R}_\text{cr}\approx100$ and a symmetric pair of stable limit cycles emerge in what appears to be a saddle-node infinite period (SNIPER) bifurcation~\citep{strogatz}. 
Although the bifurcation takes place at a relatively short distance from the spiral inlet, 
the effects of the initial point attractor persist even after the bifurcation due to a saddle-node ghost~\citep{strogatz}, leading to slow migration and further clustering of particles near the inner wall along the horizontal centreline~(see panel L in figure~\ref{Fig: BIT}(c)). We note that since the radius of curvature value at the inlet of the spirals in figure~\ref{Fig: stable limit single} are just after the SNIPER bifurcation, the dynamical origin of particle aggregation in figure~\ref{Fig: stable limit single} is also due to similar effects. After the bifurcation takes place, as shown in panel M, the pre-focused cluster of particles migrates towards the saddle point near the outer wall along its stable manifold (horizontal direction).  
Then, the pre-focused cluster of particles become more separated along the unstable manifold (vertical direction) of the saddle point as they move away from the saddle point~(see panel N). After leaving the saddle point along its unstable manifold, the particles re-cluster within a narrow range of phases along one of the pair of vertically symmetric limit cycles, similar to the behaviour observed in figure~\ref{Fig: stable limit single}. 

\subsection{Spiral ducts with bifurcations at multiple radii of curvature}\label{sec: spiral bif multiple}

\begin{figure}
\centering
\includegraphics[width=0.9\columnwidth]{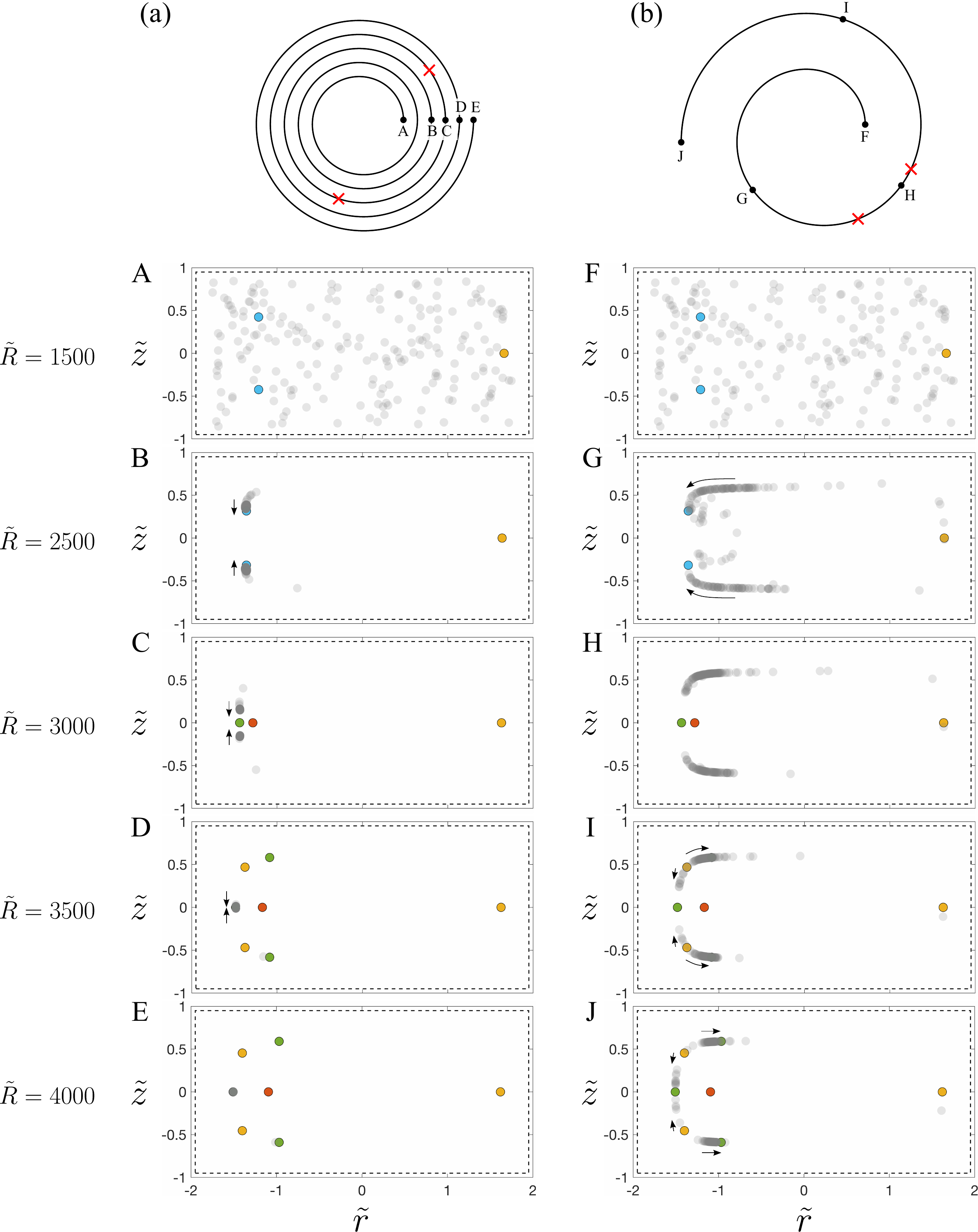}
\caption{Complex tipping phenomena with both bifurcation-induced and rate-induced effects for a flow Reynolds number of $\text{Re}=65$, in an out-spiral duct with $\tilde{R}_{\text{start}}=1500$, $\tilde{R}_{\text{end}}=4000$ and a rectangular $2\times1$ cross-section, and particles of radius $\tilde{a}=0.05$. Column (a): $N_{\text{turns}}=5$ (see Supplemental Video 3); all particles focus to the stable node on the horizontal centreline near the inner wall. Column (b): $N_{\text{turns}}=1.5$ (see Supplemental Video 4); particles focus to all three stable nodes with majority focusing near the pair of off-centred stable nodes. The difference is due to the multiple bifurcations in particle equilibria near $\tilde{R}_{cr}\approx 2850$ and $3050$ together with rate-induced effects from the different number of turns. Snapshots of the cross-section are shown at (A,F) $\tilde{R}=1500$, (B,G) $2500$, (C,H) $3000$, (D,I) $3500$ and (E,J) $4000$. The cross-sectional images show the particle equilibria as coloured circles with unstable nodes in red~(\protect\Mred), stable nodes/spirals (point attractors) in green~(\protect\Mgreen)/cyan~(\protect\Mcyan) and saddle points in yellow~(\protect\Myellow). The particle locations (grey circles -~\protect\Mgrey) and their motion (black arrows) are also shown. If the centre of a particle lies on the dashed rectangle 
it will touch at least one wall of the duct. Red cross~({\color{red}{$\boldsymbol{\times}$}}) on spirals denote the location of the bifurcations. 
}
\label{Fig: BIT_RIT}
\end{figure}

Due to rich bifurcations taking place in particle equilibria as a function of the radius of curvature, spiral ducts can cover a range of radii of curvature where more than one bifurcation takes place. This gives rise to intricate tipping phenomena that may involve a combination of bifurcation-induced, rate-induced and/or phase-induced effects.

Figure~\ref{Fig: BIT_RIT} shows an example of complex tipping phenomena arising from a combination of bifurcation-induced and rate-induced effects. We consider two out-spiral ducts with matching radius of curvature at each end but a different number of turns. The particle attractors associated with the starting radius of curvature consists of a vertically symmetric pair of stable spiral attractors. These point attractors undergo a sequence of bifurcations (a saddle-node and a supercritical pitchfork as shown in the inset of figure~\ref{Fig: bifurcations}(a)) along each spiral duct near $\tilde{R}_{cr}\approx 2850$ leading to the formation of a stable node attractor and an unstable node on the horizontal centreline of the duct cross-section; see figure~\ref{Fig: BIT_RIT}(C,H). Further along the spiral, a vertically symmetric pair of saddle-node bifurcations take place off the horizontal centreline near $\tilde{R}_{cr}\approx 3050$ leading to the birth of two additional stable node attractors. For the spiral duct with more turns (figure~\ref{Fig: BIT_RIT}, column (a)), the particles initially focus to the stable spiral attractor pair and then closely track the moving attractors. As these point attractors undergo bifurcations, the closely following particles focus to the stable node attractor on the horizontal centreline~(see also Supplemental Video 3). In the spiral duct with less turns (figure~\ref{Fig: BIT_RIT}, column (b)), the particles are unable to closely track the moving stable spiral attractor pair. The lagging particles further slow down near the location of forthcoming off-centred saddle-node bifurcations. The critical slowing prior to a saddle-node bifurcation is due to the presence of a saddle-node ghost and follows a square-root scaling law~\citep{strogatz}. After the saddle-node bifurcation takes place, the majority of the slowed particles are attracted towards the pair of newly formed off-centred stable node attractors~(see also Supplemental Video 4). Independently, the tipping phenomena arising in each case 
are due to bifurcation-induced effects. However, if we compare the two spiral ducts whose only difference is the number of turns, we see a rate-induced effect where, depending on the rate at which the radius of curvature changes, the particles ultimately focus to different point attractors of the system. Thus, the tipping observed in figure~\ref{Fig: BIT_RIT} is a combination of bifurcation-induced and rate-induced effects.

\begin{figure}
\centering
\includegraphics[width=\columnwidth]{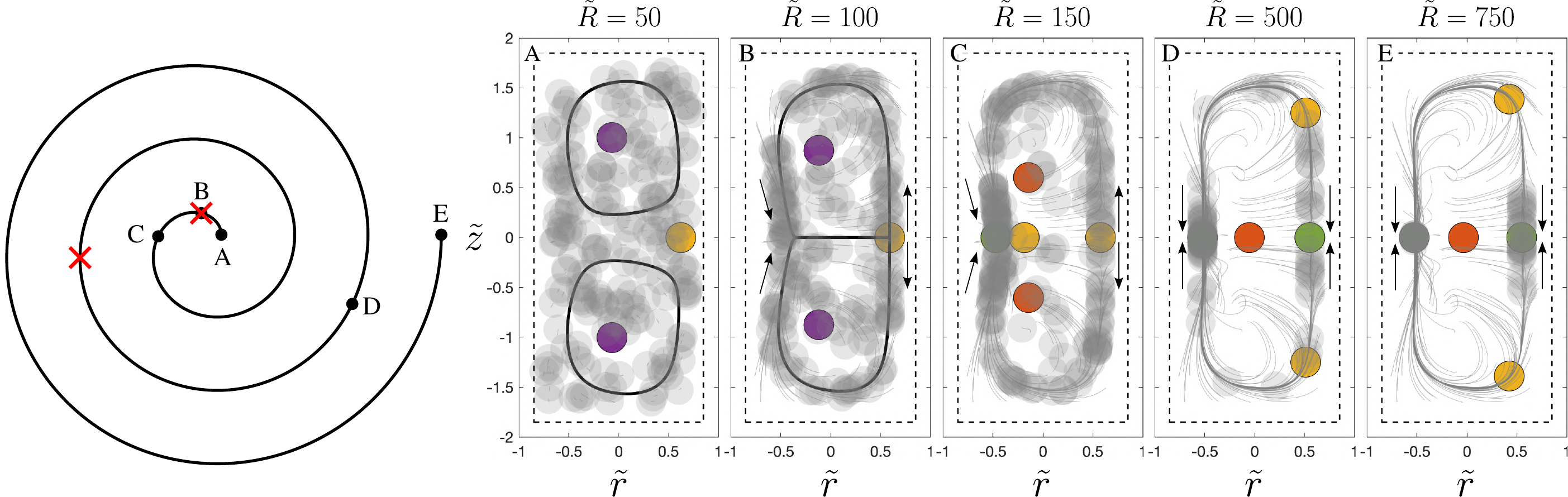}
\caption{Complex tipping phenomena with bifurcation-induced and phase-induced effects for a flow Reynolds number of $\text{Re}=25$ in an out-spiral duct with $\tilde{R}_{\text{start}}=50$, $\tilde{R}_{\text{end}}=750$, a rectangular $1\times2$ cross-section, $N_{\text{turns}}=3$, and particles of radius $\tilde{a}=0.15$. 
These are governed by bifurcations of particle equilibria at $\tilde{R}_{cr}\approx100$ and $400$ as well as the phase of the particles on the limit cycle attractors. Snapshots of the cross-section are shown at (A) $\tilde{R}=50$, (B) $100$, (C) $150$, (D) $500$ and (E) $750$. The cross-sectional images show the particle equilibria as coloured circles with unstable nodes in red~(\protect\Mred), stable nodes (point attractors) in green~(\protect\Mgreen), saddle points in yellow~(\protect\Myellow) and unstable spirals in purple~(\protect\Mpurple). Stable limit cycles are shown as black curves. The particle locations (grey circles -~\protect\Mgrey), their motion (black arrows) and trajectories (grey curves) are also shown. If the centre of a particle lies on the dashed rectangle it will touch at least one wall of the duct. Red cross~({\color{red}{$\boldsymbol{\times}$}}) on spirals denote the location of the bifurcations. 
}
\label{Fig: BIT_RIT_1x2}
\end{figure}

Another example of complex tipping phenomena that involves limit cycle attractors is shown in figure~\ref{Fig: BIT_RIT_1x2} for an out-spiral duct with a $1\times2$ rectangular cross-section containing particles of size $\tilde{a}=0.15$. A vertically symmetric pair of limit cycles exists at the starting radius of curvature. As the spiral is traversed, the limit cycle attractor pair undergoes a SNIPER bifurcation near $\tilde{R}_{cr}\approx100$ resulting in the emergence of a stable node attractor on the horizontal centreline near the inner wall. Further along the spiral near $\tilde{R}_{cr}\approx400$, a subcritical pitchfork bifurcation takes place on the horizontal centreline near the outer wall resulting in the emergence of another stable node attractor. Thus, the pair of limit cycle attractors at the inlet are replaced with two stable node point attractors at the outlet of the spiral duct. Randomly distributed particles in the $1\times2$ cross-section of this spiral initially focus on the limit cycles. However, since the period of this limit cycle diverges as the SNIPER bifurcation is approached (due to the square-root singularity arising from the saddle-node ghost~\citep{strogatz}), many particles accumulate on phases of the limit cycles near the inner wall while few particles are on the phases of the limit cycles located near the outer wall (see panel B of figure~\ref{Fig: BIT_RIT_1x2}). After the SNIPER bifurcation, the accumulated particles near the inner wall focus to the newly emerged stable node attractor and more particles approach this stable point along the slow-manifold. However, after the subcritical pitchfork bifurcation near the outer wall, a new stable node attractor emerges there too. Hence, some of the particles that were initially on phases of the limit cycle near the outer wall, fall in the basin of attraction of the stable node attractor near the outer wall and focus to this point attractor. Hence, we observe a complex tipping behaviour where in addition to the bifurcations, the phase of particles on each limit cycle determines the stable node attractor they will focus towards.


\section{Effects of flow rate}\label{sec: flow rate}

\begin{figure}
\centering
\includegraphics[width=\columnwidth]{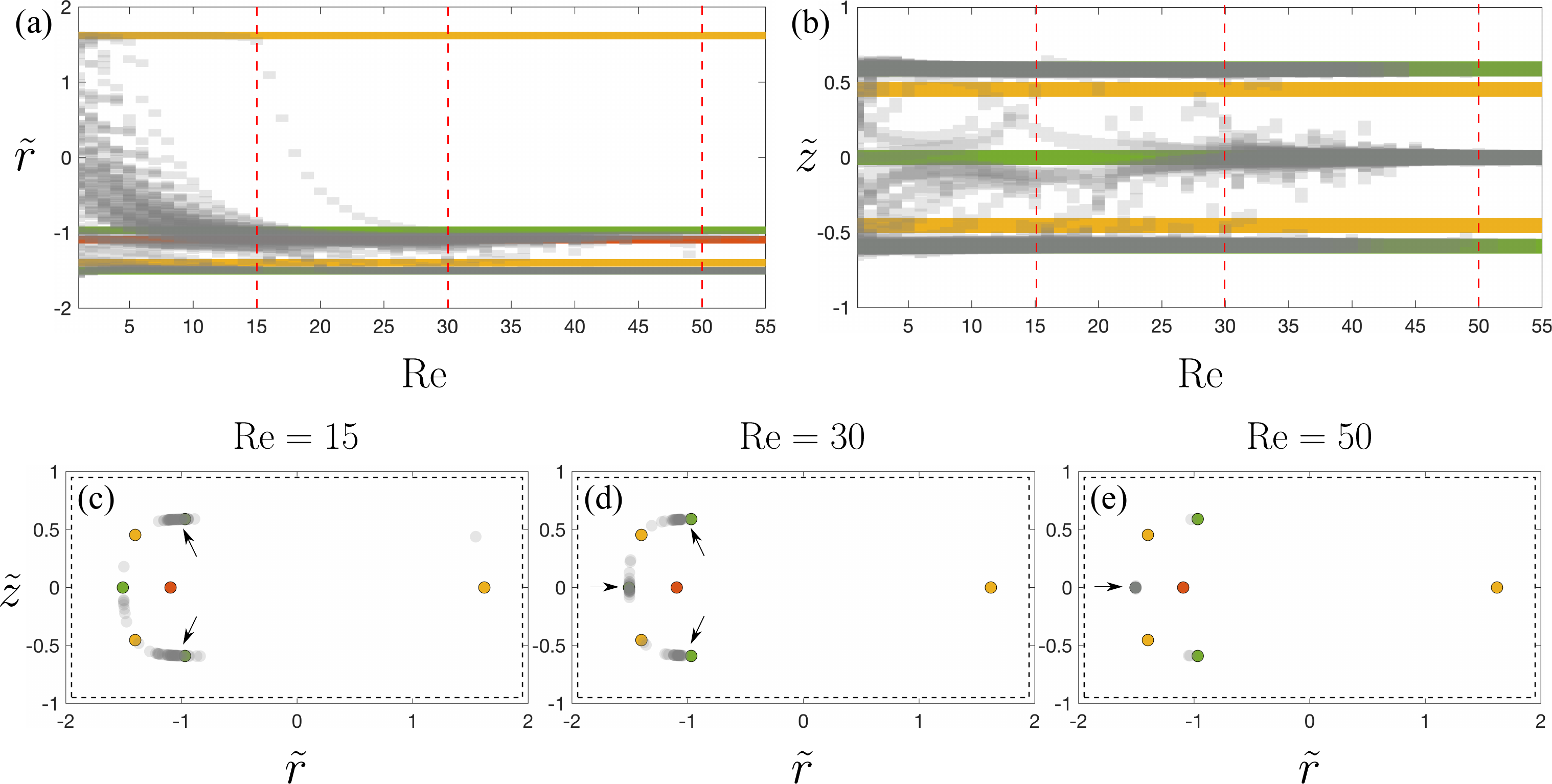}
\caption{Effect of flow rate on tipping phenomena. In an out-spiral duct with $\tilde{R}_{\text{start}}=1500$, $\tilde{R}_{\text{end}}=4000$, $N_{\text{turns}}=5$ and a rectangular $2\times1$ cross-section, particles of radius $\tilde{a}=0.05$ focus to attractors with (a) radial coordinate $\tilde{r}$ and (b) vertical coordinate $\tilde{z}$ in the final cross-section which depend on the flow Reynolds number, $\text{Re}$, as shown. Snapshots of the final cross-section are shown for (c) $\text{Re}=15$, (d) $30$ and (e) $50$ and these values of $\text{Re}$ are marked by the dashed red lines in (a) and (b).
The coloured bars in (a)--(b) and coloured circles in (c)--(e) indicate locations of particle equilibria at the end radius of curvature with unstable nodes in red~(\protect\Mred), stable nodes (point attractors) in green~(\protect\Mgreen) and saddle points in yellow~(\protect\Myellow), while the grey bars and grey circles~(\protect\Mgrey) show particle positions at the end of the spiral duct. The black arrows in (c)--(e) show the attractors to which particles are focused while the dashed rectangle marks the location of particle centres such that they will touch at least one wall of the duct.}
\label{Fig: Effects_Re}
\end{figure}

The mathematical model presented in section~\ref{sec: model} is valid for small particle Reynolds number i.e. $\text{Re}_p=\text{Re}\,\tilde{a}^2/4 \ll 1$, which also restricts our flow Reynolds number $\text{Re}$ and hence the flow rate, based on the particle size $\tilde{a}$. 
Additionally, we require that the Dean number is at most $K=\mathrm{Re}^2/\tilde{R}=O(10)$, which effectively restricts the flow rate depending on the (minimum) radius of curvature.
In principle, we can explore the dynamics up to Dean numbers $K=O(100)$ using the improved theoretical model of \citet{harding_stokes_2023}, but this is beyond the scope of the present work.
For the range of particle sizes and radii of curvature considered in this paper, an upper limit of $\text{Re}=100$ is reasonable~\citep{harding_stokes_bertozzi_2019,Valani2022SIADS}. 
Within the framework of our low flow-rate model, the cross-sectional migration velocities are proportional to $U_m^2$ while the axial flow scales with $U_m$.
Equivalently, given fixed fluid properties and duct geometry, each scale with $\mathrm{Re}^2$ and $\mathrm{Re}$, respectively.
Hence, for a fixed spiral duct geometry, we expect slower particle focusing at lower flow rates (i.e. smaller $\text{Re}$) compared to higher flow rates (i.e. larger $\text{Re}$). 
For example, upon halving the flow rate $U_m$, a particle will take twice as long to traverse the spiral but migration velocities will reduce by a factor of one quarter, with the net result that (roughly) half the migration distance is achieved per unit arc-length along the spiral curve.

We demonstrate dependence on flow rate by revisiting the example of complex tipping phenomena shown in figure~\ref{Fig: BIT_RIT} for a range of $\text{Re}$ values, as shown in figure~\ref{Fig: Effects_Re}. At a low value of $\text{Re}=15$ (figure~\ref{Fig: Effects_Re}(c)), the cross-sectional migration is relatively slow and particles are unable to closely track the moving stable spiral attractors. This results in clustering of particles near the pair of off-centred stable node attractors at the end of the spiral duct. At an intermediate Reynolds number of $\text{Re}=30$ (figure~\ref{Fig: Effects_Re}(d)), some particles are able to track the moving attractors more closely while others cannot, and this results in a division of particle focusing at the end of the spiral duct; some particles focus to the stable node attractor on the centreline while others, caught on the opposite side of a saddle-node ghost, focus to the pair of symmetric off-centred stable node attractors. At a larger value of $\text{Re}=50$, cross-sectional migration is faster and all particles are able to closely track the point attractors and focus to the stable node attractor on the centreline. Thus, we see that rate-induced tipping effects can also arise from variations in the flow rate i.e. the flow Reynolds number $\text{Re}$, in addition to variations in $N_{\text{turns}}$ as was shown in figure~\ref{Fig: BIT_RIT}.
We note that the flow rate is a much more immediately manipulable control parameter for experiments compared to the number of turns (which requires the creation of a new device for each adjustment).


\begin{figure}
\centering
\includegraphics[width=\columnwidth]{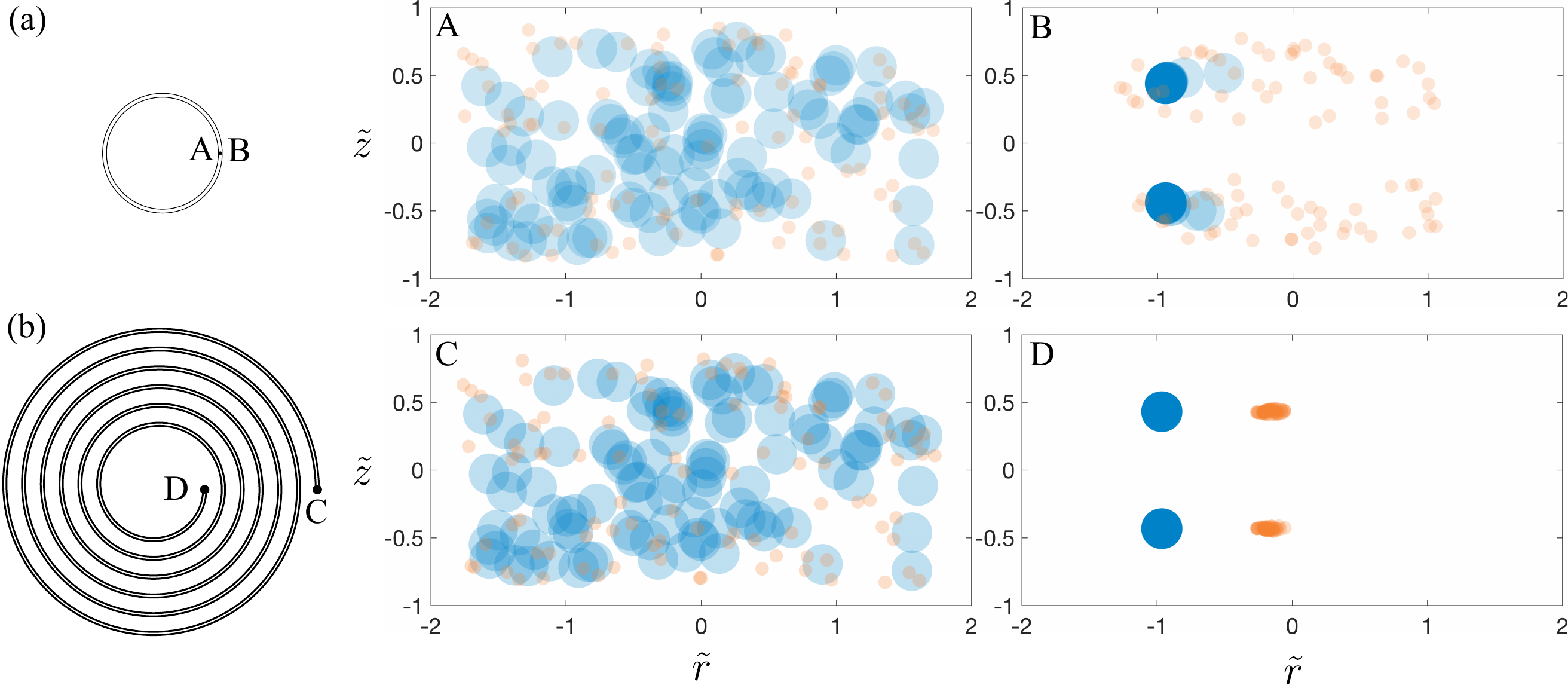}
\caption{Comparison of circular and spiral ducts for particle separation. Row (a): in a circular duct with a $2\times1$ rectangular cross-section and a constant radius of curvature $\tilde{R}=60$, (A) initially random distributions of particles of radii $\tilde{a}=0.05$ (orange) and $0.15$ (blue) are unable to focus to their respective stable spiral attractors at (B) the end of one full turn. Row~(b): conversely, in an in-spiral duct with $\tilde{R}_{\text{start}}=200$, $\tilde{R}_{\text{end}}=60$ and $N_{\text{turns}}=6$, having the same cross-section and particle sizes, (C) the same initially random distribution of particles in the cross-section are able to focus and separate well at (D) the end of the spiral duct. The flow Reynolds number is fixed to $\text{Re}=100$.}
\label{Fig: PS1}
\end{figure}

\section{Non-equilibrium particle separation}\label{sec: particle separation}

In this section, we discuss the implications of particle focusing dynamics and tipping phenomena for separation of particles by size. We restrict ourselves to explore separation of particles with two different sizes noting that the mechanisms described here may be extended to separate more than two particle sizes.

When no bifurcations are present along a spiral duct for the different particle sizes to be separated, a spiral duct provides an advantage over a circular duct at relatively small bend radii where one turn of a circular duct might be insufficient to achieve complete focusing of all particles. 
An example is shown in figure~\ref{Fig: PS1}. Particles of size $\tilde{a}=0.05$ (orange) and $0.15$ (blue) in a circular duct with radius of curvature $\tilde{R}=60$ have a vertically symmetric pair of stable spiral attractors in the cross-section that are spatially separated along the width of the duct. However, due to insufficient length afforded by one turn, not all particles are able to completely focus to their respective attractors~(figure~\ref{Fig: PS1}(a)); in particular the smaller particles are far from focused and separation by size has not been achieved. In contrast, if one considers the same two particle sizes in an in-spiral duct that starts with a larger radius of curvature, $\tilde R_\text{start}=200$, and ends at the same radius of curvature as the circular duct, then focusing of all particles to their respective point attractors is achieved resulting in complete separation of the two particle sizes (figure~\ref{Fig: PS1}(b)).

\begin{figure}
\centering
\includegraphics[width=1\columnwidth]{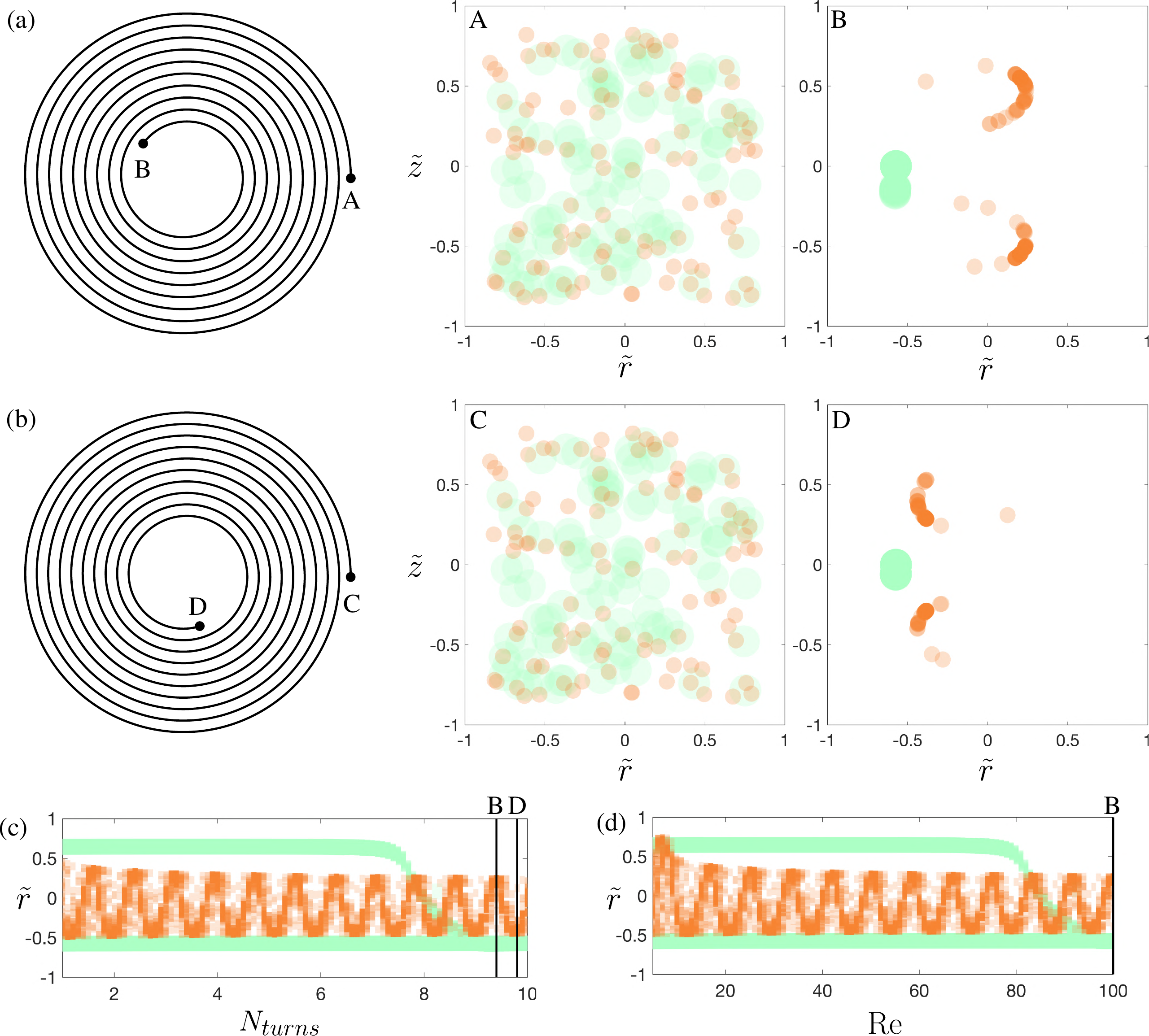}
\caption{Particle separation in an in-spiral duct with a square cross-section, $\tilde{R}_{\text{start}}=1250$, $\tilde{R}_{\text{end}}=400$, and particles of radius $\tilde{a}=0.05$ (orange) and $0.10$ (green). The larger particles have a stable node attractor and the smaller particles have a pair of symmetric limit cycle attractors. 
Row (a): after $N_{\text{turns}}=9.4$ the smaller particles are sufficiently focused on the limit cycle within a range of phase angle that separates them in the radial direction from the almost focused larger particles, such that good separation by size is achievable. Row (b): after $N_{\text{turns}}=9.8$ the smaller particles are focused on the limit cycle within a narrower range of phase angle but their radial proximity to the focused larger particles results in poor separation by size. For both spiral ducts, cross-sections A and C show the particle positions at the start of the spiral while cross-sections B and D show the particle positions at the end of the spiral. The flow Reynolds number is fixed to $\text{Re}=100$. (c) Radial position $\tilde{r}$ of the particles at the end of the spiral as a function of the number of turns $N_{\text{turns}}$ for $\text{Re}=100$. (d) Radial position $\tilde{r}$ of the particles at the end of the spiral as a function of $\text{Re}$ for $N_{\text{turns}}=9.4$.}
\label{Fig: PS2}
\end{figure}

In the case of two different particle sizes for which one or both have a limit cycle attractor,
the length of the spiral duct (i.e. $N_{\text{turns}}$) or the flow rate (i.e. $\text{Re}$) may be used as parameters to control the phase of the particles along the limit cycle(s) and, so, achieve separation by size. Recall figure~\ref{Fig: stable limit single} which shows particles focused to, but non-uniformly distributed on, the limit cycle such that the majority lie within a small range of the phase angle, dependent on $N_\text{turns}$. 
Exploitation of particle clustering on limit cycles is shown in figure~\ref{Fig: PS2} for particles of size $\tilde{a}=0.05$ (orange) and $0.10$ (green) suspended in a flow through an in-spiral duct with a square cross-section. The attractors for the smaller particle are a vertically symmetric pair of limit cycles throughout the spiral~(similar to figure~\ref{Fig: stable limit single}). The larger particle initially has two stable node attractors located on the horizontal centreline, one near the inner wall and the other near the outer wall; however, further along the duct, the stable node attractor near the outer wall vanishes in a subcritical pitchfork bifurcation and only the stable node attractor near the inner wall persists through to the final cross-section of the spiral. For $N_{\text{turns}}=9.4$ shown in figure~\ref{Fig: PS2}(a), the larger particles initially focus to both the centreline point attractors but, by the end of the spiral, all larger particles focus to the sole stable node attractor near the inner wall. The smaller particles are clustered on a phase of the limit cycle which places them nearest to the outer wall at the end of the spiral. This result provides good separation of the two particles with respect to the final radial location $\tilde{r}$. However, on slightly increasing the number of turns to $N_\text{turns}=9.8$ and, hence, the length of the spiral, there is a near 180$^\circ$ phase shift for the smaller particles focused on the limit cycle, as shown in figure~\ref{Fig: PS2}(b), resulting in poor radial separation between the different sized particles. Figure~\ref{Fig: PS2}(c) shows the radial position $\tilde r$ of the particles (larger particles in green, smaller in orange) as a function of $N_\text{turns}$ from which a suitable number of turns for good particle separation can be selected; panels B and D are marked by vertical black lines. Figure~\ref{Fig: PS2}(d) shows $\tilde{r}$ for the particles as a function of Reynolds number $\text{Re}$, where $N_\text{turns}=9.4$; panel B is marked by a vertical black line. It is clear that $\text{Re}$ (or flow rate) might also be chosen to achieve a good separation of the particles by size.

An alternative method of manipulating the phase of particles on a limit cycle, known as Dean Flow Fractionation~\citep{Bhagat2008,Hou2013}, is exploited in inertial microfluidic experiments. In this method, the spiral duct has two inlets, one near the inner wall consisting of a sample containing particles (typically a mix of two distinct sizes), and the other near the outer wall through which a particle-free `sheath' flow is introduced. This, effectively, pre-focuses the particles near the inner wall. As the particles flow through the curved duct, the Dean vortices transport the smaller particles towards the outer wall along what appears to be a limit cycle attractor, while the larger particles equilibrate at a stable node attractor near the inner wall, thus achieving separation between the two particle sizes over a relatively short duct length. 
Figures~\ref{Fig: stable limit single} and \ref{Fig: PS2} above illustrate that phase angle on a limit cycle can potentially be manipulated without the need for a sheath flow through a second inlet. 
Specifically, the phase of smaller particles on a limit cycle may be controlled in two ways: (i) using the number of turns (see figure~\ref{Fig: PS2}(c)) or (ii) the flow rate (see figure~\ref{Fig: PS2}(d)). 

\begin{figure}
\centering
\includegraphics[width=\columnwidth]{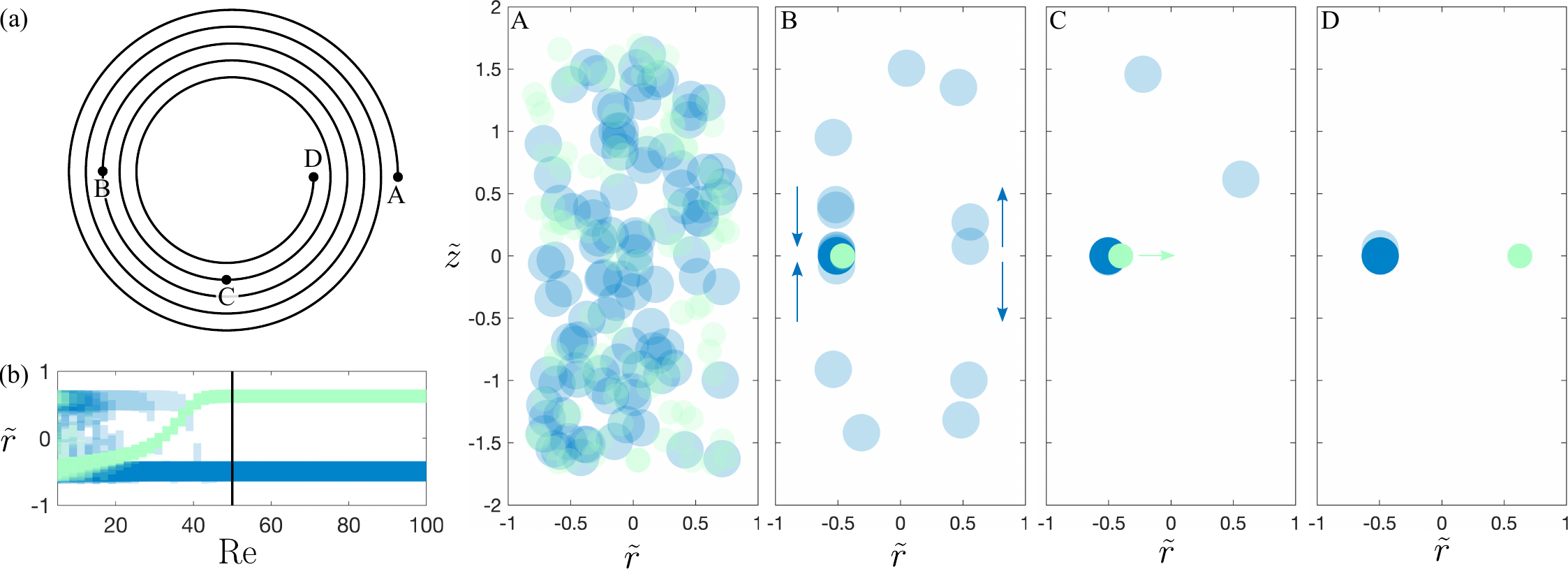}
\caption{Particle separation in (a) an in-spiral duct with 
$\tilde{R}_{\text{start}}=400$, $\tilde{R}_{\text{end}}=200$, $N_{\text{turns}}=5$ and a rectangular $1\times2$ cross-section. (A) Randomly distributed small particles (green) of radius $\tilde{a}=0.10$ and 
larger particles (blue) of radius $\tilde{a}=0.15$ initially focus to (B) stable nodes near the inside wall after which separation occurs via B-tipping with loss of the stable node attractor for the smaller particles such that (C,D) the smaller particles transiently focus to a saddle point near the outer wall while the bigger particles remain focused at the stable node attractor near the inner wall. Snapshots A--D are shown at $\tilde{R}=400, 300, 250, 200$, respectively. The flow Reynolds number is fixed to $\text{Re}=50$. (b) Variation in the radial position $\tilde{r}$  of particles in the final cross-section as a function of the flow Reynolds number $\text{Re}$; the black vertical line indicates the value $\text{Re}=50$.}
\label{Fig: PS3}
\end{figure}

We now recall 
figure~\ref{Fig: BIT}, which showed bifurcations such that a particle attractor became unstable resulting in the tipping of focused particles away from the location of that attractor. 
This might be exploited to give a new type of transient particle separation within spiral duct geometries.
Consider a spiral duct containing particles of two distinct sizes. Suppose that the particles first focus to two attractors, one for each size, after which one of these particle sizes is affected by B-tipping within the spiral so that its attractors are lost and, perhaps, others are born.
Those particles affected by the B-tipping will then undergo migration towards other attractors while those not affected by the B-tipping remain focused.
This could be undesirable if it occurs near the end of a spiral duct and results in de-focusing within the final cross-section.
However, if the particles affected by B-tipping become sufficiently tightly focused prior to the bifurcation, then they will remain clustered for some time post-bifurcation. 
During this time, the transient motion of the pre-focused cluster might result in a large separation from the particles of different size that remain focused such that particle separation is achieved.
An example was first shown in \citet{ValaniPoF2023} in which a large separation is achieved between two different sized particles in a spiral duct with square cross-section by utilising bifurcations in this way. A similar mechanism of particle separation can also be achieved within taller rectangular ducts~(see figure~\ref{Fig: PS3}) and is explained below. 

For the chosen in-spiral in figure~\ref{Fig: PS3}, the larger particles (blue) have a stable node attractor on the horizontal centreline near the inner wall which persists throughout the spiral. 
All of the larger particles focus to this attractor as a consequence. 
Initially the smaller particles also have a stable node attractor on the horizontal centreline near the inner wall to which they focus. However, a SNIPER bifurcation then occurs and the stable node attractor for the smaller particles is replaced by a pair of limit cycle attractors, following which the pre-focused cluster of smaller particles migrates along the stable manifold of a saddle-point located on the centreline near the outer wall and temporarily focus around the saddle point. 
It takes some time before the particles disperse away from the saddle point (vertically, in both directions) because the eigenvalue associated with the unstable manifold is an order of magnitude smaller than that associated with the stable manifold~\citep{ValaniDSTA2021,Valani2022SIADS}.
If the smaller particles reach the end of the spiral before dispersion can take place, then the different sized particles will be successfully separated, as in panel D of figure~\ref{Fig: PS3}. 
If the smaller particles are given time to disperse they will ultimately become attracted onto a limit cycle pair.
We note that such particle dynamics have been observed in the experiments of \citet{Cruz2021} suggesting bifurcations might play a role in their observed behaviours. 

The robustness of this separation mechanism is demonstrated in figure~\ref{Fig: PS3}(b) which shows the radial positions $\tilde{r}$ of the two different sized particles at the end of the spiral duct as a function of the Reynolds number $\text{Re}$. It can be seen that for small $\text{Re}$ ($\lesssim 20$), the particles in the spiral duct do not completely focus (especially the larger ones). For $20\lesssim \text{Re} \lesssim 40$, we observe that both particle sizes are focused but not well separated in the radial ($\tilde{r}$) direction. For $40\lesssim\text{Re}\lesssim 100$, we find that the particles are well focused and separated in the radial direction. Eventually, for even larger $\text{Re}$ (not shown), the transiently focused smaller particles near the saddle point start to disperse along the unstable manifold of the saddle and migrate towards the inner wall along the limit cycle attractors, resulting in poor separation. Hence, we have a large intermediate range of Reynolds numbers in which the separation is clean and insensitive to $\text{Re}$.

\begin{figure}
\centering
\includegraphics[width=\columnwidth]{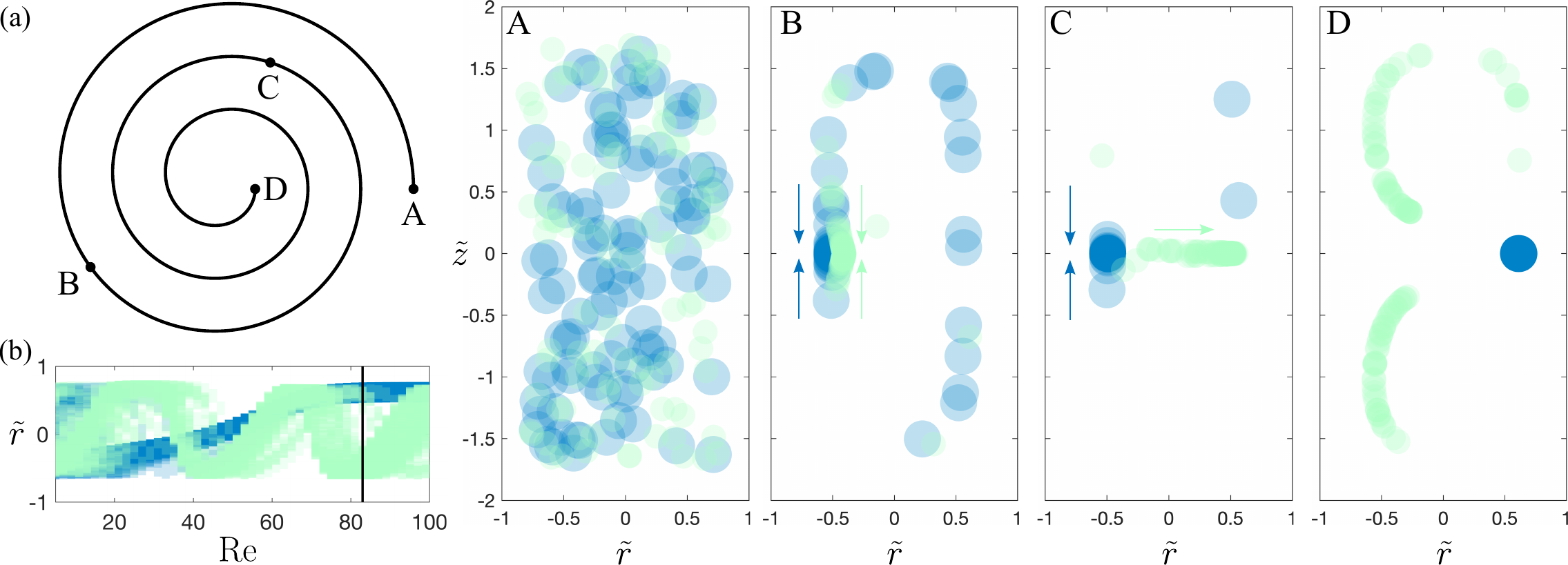}
\caption{
Particle separation in (a) an in-spiral duct with 
$\tilde{R}_{\text{start}}=300$, $\tilde{R}_{\text{end}}=50$, $N_{\text{turns}}=3$ and a rectangular $1\times2$ cross-section. (A) Randomly distributed small particles (green) of radius $\tilde{a}=0.10$ and 
larger particles (blue) of radius $\tilde{a}=0.15$ initially focus to (B) their respective stable nodes near the inside wall after which separation occurs via B-tipping with loss of the stable node attractor for both particles such that (C,D) the smaller particles focus on limit cycles while the larger particles transiently focus to a saddle point near the outer wall. Snapshots A--D are shown at $\tilde{R}=300, 250, 200, 50$, respectively. The flow Reynolds number is fixed to $\text{Re}=83$. (b) Variation in the radial position $\tilde{r}$  of particles in the final cross-section as a function of the flow Reynolds number $\text{Re}$; the black vertical line indicates the value $\text{Re}=83$.
}
\label{Fig: PS3_2}
\end{figure}

\begin{figure}
\centering
\includegraphics[width=\columnwidth]{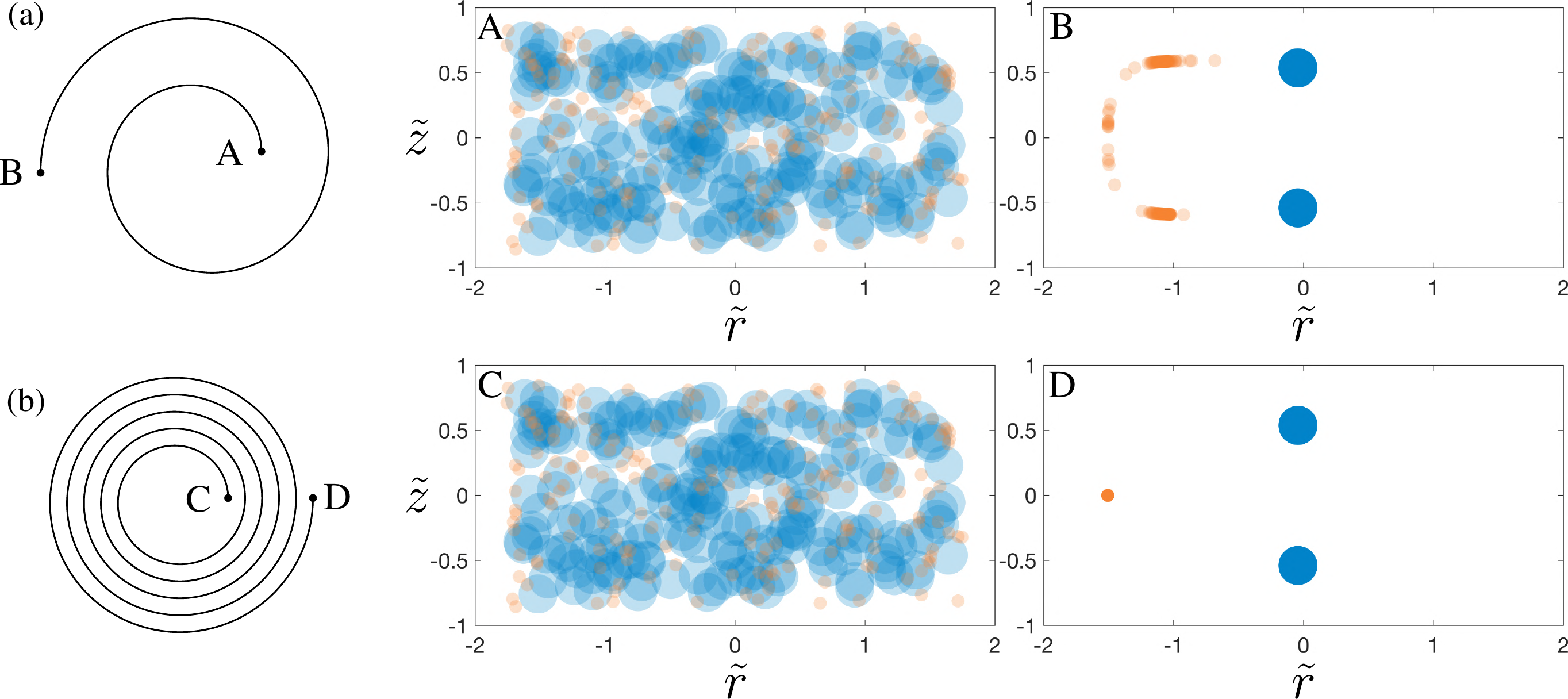}
\caption{
In out-spiral ducts with $\tilde{R}_{\text{start}}=1500$, $\tilde{R}_{\text{end}}=4000$ and a rectangular $2\times1$ cross-section, small particles of radius $0.05$ (orange) undergo different tipping phenomena (see figure~\ref{Fig: BIT_RIT}) for row (a) $N_{\text{turns}}=1.5$ and row (b) $N_{\text{turns}}=5$, while larger particles of radius $0.15$ (blue) have persistent point attractors. Snapshots showing particle positions are shown at the (A,C) start and (B,D) end of the spiral ducts. Here the flow Reynolds number is fixed to $\text{Re}=50$.}
\label{Fig: PS4}
\end{figure}


One can also exploit B-tipping to achieve particle separation between two distinct particle sizes even when no point attractors exist for either particle size at the ending radius of curvature. 
For example, figure~\ref{Fig: PS3_2} shows an in-spiral duct with a $1\times2$ rectangular cross-section in which small particles (green) initially migrate towards a stable node attractor located on the horizontal centreline near the inner wall. 
However, due to an immediate SNIPER bifurcation, the stable node attractor vanishes and a symmetric pair of limit cycle attractors are born. 
The smaller particles pre-focus near the inner wall~(panel B), then migrate along the stable manifold of the saddle point located on the horizontal centreline near the outer wall~(panel C) and then ultimately onto the 
attracting limit cycles while remaining clustered in a narrow range of phase angles~(panel D), similar to the mechanism in figures~\ref{Fig: stable limit single} and \ref{Fig: PS2}. The larger particles (blue) also initially focus at a stable-node point attractor on the horizontal centreline near the inner wall. However, a SNIPER bifurcation again causes this point attractor to vanish and the pre-focused cluster of larger particles migrates along the horizontal centreline towards the outer wall, i.e. along the stable manifold of a saddle point near the outer wall~(panels C and D). 
If, at the end of the spiral duct, the larger particles have not yet dispersed from their saddle point and the smaller particles are 
on the portion of the limit cycle closest to the inner wall, then a good separation between the two particle sizes is achieved (as in panel D of figure~\ref{Fig: PS3_2}).
Although a similar SNIPER bifurcation took place along the spiral duct for both particle sizes, it occurred earlier along the spiral for the smaller particles compared to the larger particles, allowing the smaller particles to settle onto limit cycle attractors before the larger particles temporarily focus near the outer wall saddle. 
We conclude that the mechanism depicted in figure~\ref{Fig: PS3_2} is a complete non-equilibrium particle separation mechanism because neither of the two particle sizes are focused at point attractors and will exhibit significant motion if the spiral duct was extended. 
Figure~\ref{Fig: PS3_2}(b) shows how this mechanism is affected by changes to the Reynolds number. For $\text{Re} \gtrsim 75$ the larger particle remains near the saddle point at the outer wall at the end of the spiral while the focusing position of the cluster of smaller particles oscillates along the limit cycle attractor. This provides multiple opportunities to `tune' this outcome in a practical setting by manipulating the flow rate. However, we note that this will not continue indefinitely for increasing $\text{Re}$ since the transient focusing of larger particles at a saddle point will cease for large enough $\text{Re}$ and the presented separation mechanism will be ineffective.

We observed in figure~\ref{Fig: RIT} that purely rate-induced effects can cause particles to end up in a basin of attraction different from the one they start in. Since such effects occur for a very small proportion of initial particle positions in the cross-section, we do not expect them to have a significant impact on particle separation by size. However, we observed in figure~\ref{Fig: BIT_RIT} that rate-induced effects become more significant when combined with bifurcation-induced effects, and this can have implications on particle separation. 
Figure~\ref{Fig: PS4} illustrates this for a suspension of particles of two different sizes flowing along the two in-spiral ducts of figure~\ref{Fig: BIT_RIT}. In both spirals, the larger particles (blue) focus to a pair of point attractors near the centre of the cross-section. However, the point attractors to which the smaller particles (orange) focus depend on the number of turns (as previously discussed); specifically they focus to the inner most (left) stable node attractor in the spiral having more turns and to the pair of off-centred stable node attractors in the spiral having fewer turns. Hence, the number of turns in the spiral influences the tipping phenomena of the smaller particles and, in turn, the degree of separation between the two particle sizes which is achieved. 

\section{Conclusion}\label{sec: conclusions}

We have explored the migration dynamics of particles suspended in fluid flow through Archimedean-like spiral ducts at low flow rates. This work has revealed the richness in particle dynamics and focusing behaviours due to the changing nature of particle attractors within the duct cross-section as the radius of curvature changes along the spiral duct. We have made connections between these behaviours and various types of tipping phenomena which have been explored in other systems.

When no bifurcations occur throughout the duct, the primary advantage provided by spiral duct designs over circular ducts is the increased duct length that can facilitate more complete focusing of particles to their respective attractors. When the cross-sectional attractors are $1$D limit cycles, then their evolving size and period throughout the spiral can be exploited to cluster a majority of the particles in a narrow range of phases along each limit cycle. When multiple radially-separated point attractors are present in the cross-section of a circular duct, a particle's initial location with respect to the basins of attraction determines the attractor to which it will focus. However, in a spiral duct, the small fraction of particles which, at the start of the spiral, are near the boundaries which separate the basins of attraction of the attractors for that cross-section 
can, further along the spiral and depending on their migration rate, see a change in their basin of attraction  and, so, a change in the attractor to which they focus and thus exhibit rate-tipping or R-tipping. 
The occurrence of R-tipping is controlled by the rate of change of the radius of curvature from the perspective of the particle, which can be controlled geometrically via the number of turns (given fixed starting and ending radii of curvature), or physically/practically via the flow rate.

Since rich bifurcations in particle equilibria can take place with changes in the radius of curvature of the duct, they are present in many spiral duct designs. Bifurcations have a significant effect on particle focusing behaviour. When particles focus to an attractor which suddenly becomes unstable (or vanishes) due to a bifurcation, it results in bifurcation-tipping, or B-tipping, and leads to further migration of affected particles to a new/different attractor of the system. As with R-tipping, B-tipping can be controlled by varying either the number of turns or the flow rate, furthermore, it can be manipulated by varying the starting or ending radius of curvature of the spiral. 
Combinations of B-tipping and R-tipping are possible and result in complex tipping phenomena in which particles can be selectively focused by size to different attractors in the cross-section. 

We have also shown that the rich dynamical and tipping behaviours exhibited by particles can facilitate novel mechanisms of particle separation. Specifically, we have identified non-equilibrium separation regimes in which at least one particle size is not focused to a point attractor. For example, we demonstrated particle separation mechanisms where: (i) smaller particles cluster along limit cycle attractors near the outer wall while larger particles focus to a stable node attractor near the inner wall~(figure~\ref{Fig: PS2}), (ii) smaller particles transiently focus to a saddle point near the outer wall while larger particles focus to a stable node attractor near the inner wall~(figure~\ref{Fig: PS3}), and (iii) smaller particles cluster along limit cycle attractors near the inner wall while larger particles transiently focus near a saddle point near the outer wall ~(figure~\ref{Fig: PS3_2}). 
We have also shown that complex combinations of B-tipping and R-tipping can influence particle separation. For example, we have demonstrated that smaller particles can be selectively focused to multiple stable node attractors near the inner wall while larger particles focus to a vertically symmetric pair of point attractors further away from that wall~(figure~\ref{Fig: PS4}). 

By further variation of the system parameters, it may be possible to identify additional non-equilibrium particle separation mechanisms that were not considered in this paper. For example, we restricted our study to rectangular cross-sections but changing the cross-sectional shape is likely to change 
the number and type of particle attractors present in a cross-section, leading to tipping phenomena and separation mechanisms that are not realised for rectangular cross-sections. 
It is known that cusp bifurcations occur in circular ducts with trapezoidal cross-sections~\citep{HardingTrapezoid}, which immediately suggests the existence of new ways of particle separation in circular and spiral ducts with trapezoidal cross-sections. Moreover, it may also prove fruitful to analyse different curved duct geometries beyond circular and spiral designs. Recent advances in inertial microfluidics has led to innovation in curved duct geometries~\citep{D0LC00714E}, e.g.  symmetric/asymmetric sinusoidal ducts and double spiral ducts, and it would be interesting to explore the variations in particle equilibria along these ducts and the corresponding particle dynamics. 

Clearly, this work would benefit from 
experimental validation. 
Successful experimental realisation of the rich dynamical behaviours and the corresponding particle separation mechanisms we have identified could open new avenues for size-based particle separation for many different applications.


\section*{Acknowledgements}
We would like to thank Dr John Maclean, School of Computer and Mathematical Sciences, The University of Adelaide, for bringing our attention to tipping phenomena in dynamical systems. This research is supported under the Australian Research Council’s Discovery Projects funding scheme (project number DP200100834). The results were computed using supercomputing resources provided by the Phoenix HPC service at the University of Adelaide and the R\={a}poi HPC service at Victoria University of Wellington.

\section*{Declaration of interests} 
The authors report no conflicts of interest.

\begin{figure}
\centering
\includegraphics[width=0.9\columnwidth]{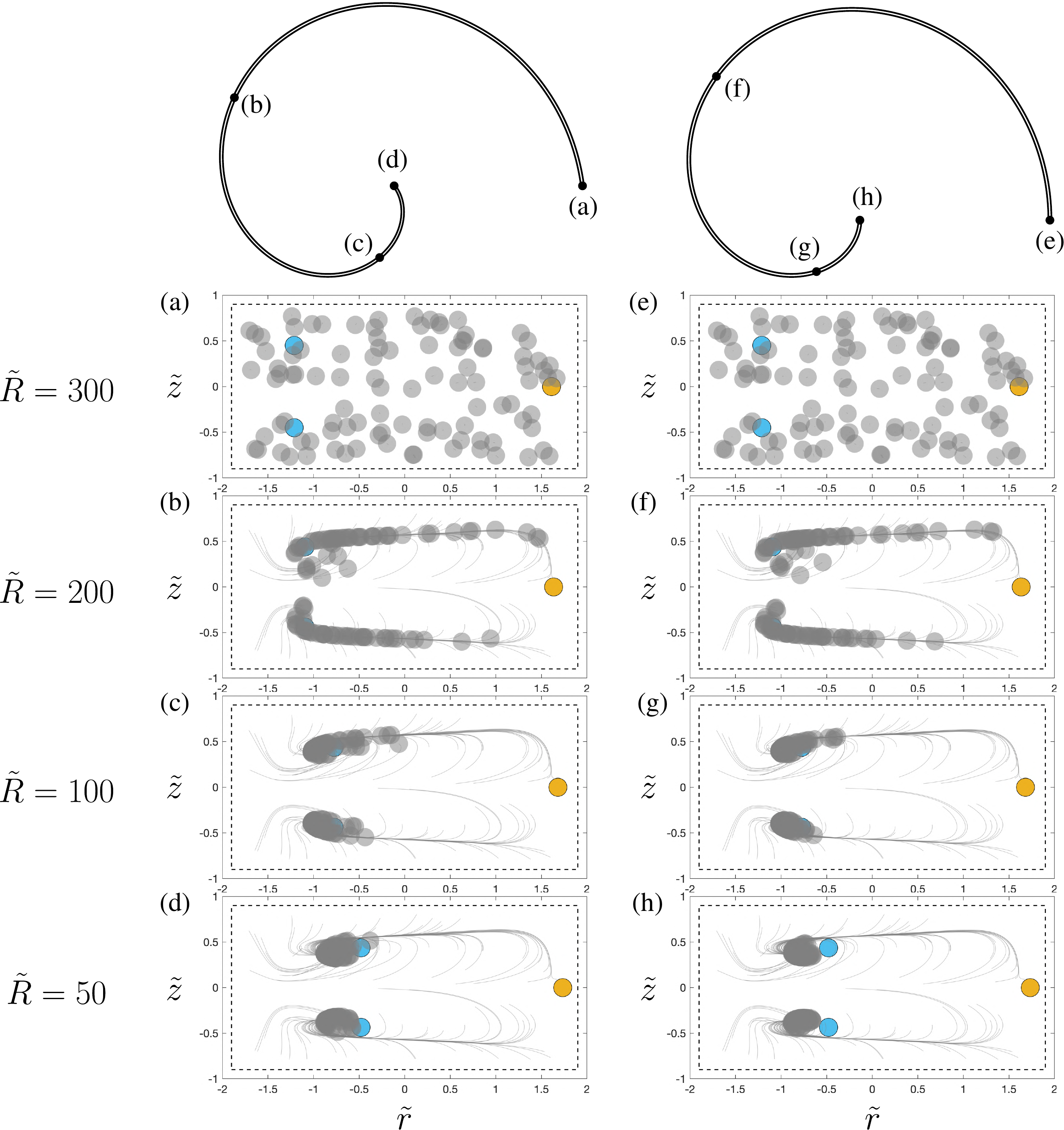}
\caption{Comparison of particle focusing dynamics in (a--d) a traditional Archimedean spiral duct and (e--h) an Archimedean-like spiral duct, both having a $2\times 1$ rectangular cross-section and parameters $\tilde R_{\text{start}}=300$, $\tilde R_{\text{end}}=50$, $N_\text{turns}=1$, $\text{Re}=75$, $\tilde{a}=0.10$. 
Snapshots of the cross-section 
are shown at $\tilde{R}=300,200,100$ and $50$. Note that $\tilde{R}$ is (a,b,c,d) the local bend radius of the Archimedean spiral centreline and (e,f,g,h) the local radius of curvature of the Archimedean-like spiral, respectively. The coloured circles denote the particle equilibria with cyan~(\protect\Mcyan) for stable spirals and yellow~(\protect\Myellow) for saddle points. The particle locations are shown as grey circles~(\protect\Mgrey) while particle trajectories are grey curves. If the centre of a particle lies on the dashed rectangle 
it will touch at least one wall of the duct. 
}
\label{Fig: compare rad vs local rad stable spiral single}
\end{figure}

\appendix
\section{}\label{A}

In this appendix we compare the particle dynamics in 
two spiral ducts with slightly different geometries: (i) an Archimedean-like spiral described by equation~\eqref{Archimedean-like spiral}, similar to those used throughout this paper, and (ii) a traditional Archimedean spiral described by equation~\eqref{Archimedean spiral}; see 
figure~\ref{Fig: compare rad vs local rad stable spiral single}, the caption of which gives the parameters which have the same values for both spirals. Recall, however, that the parameter $\tilde R$ is defined differently for the two spirals. In the case of the Archimedean spiral this denotes the local bend radius, while for the Archimedean-like spiral it denotes the local radius of curvature.

This figure shows that the slight difference in the spiral geometries makes little difference to the particle dynamics within them. The two cross-sections shown at the four different values of $\tilde R$ are very similar, both qualitatively and quantitatively. Hence, we do not expect the results presented in this paper to change appreciably should the Archimedean-like spirals be replaced by traditional Archimedean spirals. 

\bibliographystyle{jfm}
\bibliography{jfm-instructions}

\end{document}